\documentclass[10pt,nocopyrightspace,onecolumn,preprint,numbers]{sigplanconf}
\usepackage{etex}
\usepackage[usenames,dvipsnames,svgnames,table]{xcolor}
\usepackage{prooftree}
\usepackage{listings}
\usepackage{pifont}
\usepackage{multirow}
\usepackage{amsmath}
\usepackage{amsfonts}
\usepackage{amssymb}
\usepackage{xspace}
\usepackage{graphicx}
\usepackage{lastpage}
\usepackage{paralist}
\usepackage{fancyhdr}
\usepackage{wrapfig}
\usepackage{balance}
\usepackage{tikz}
\usepackage{ctable} 
\usepackage{pgfplots}
\usepackage{amsthm}
\usepackage{dsfont}
\usepackage{colonequals}
\usepackage{enumitem}
\usepackage{wasysym}
\usepackage{mathtools}
\usepackage{listings}
\usepackage{algorithm}
\usepackage[noend]{algpseudocode}
\usepackage{pifont}
\usepackage{stmaryrd}
\usepackage{comment}
\usepackage[hyphens]{url}
\usepackage{etoolbox}
\usepackage{pbox}
\usepackage{relsize}
\usepackage{csquotes}
\usepackage{scalerel}
\usepackage{setspace}
\usepackage[T1]{fontenc}
\usepackage{manfnt}
\usepackage{tabularx}
\usepackage{hhline}
\usepackage{titlesec}
\usepackage{enumitem}
\usepackage[percent]{overpic}

\titlespacing*{\section}
{0pt}{1.3ex plus .1ex minus .1ex}{1.ex plus .1ex}
\titlespacing*{\subsection}
{0pt}{1.3ex plus .1ex minus .1ex}{1.ex plus .1ex}

\DeclareFontFamily{U}{mathx}{\hyphenchar\font45}
\DeclareFontShape{U}{mathx}{m}{n}{
      <5> <6> <7> <8> <9> <10>
      <10.95> <12> <14.4> <17.28> <20.74> <24.88>
      mathx10
      }{}
\DeclareSymbolFont{mathx}{U}{mathx}{m}{n}
\DeclareFontSubstitution{U}{mathx}{m}{n}
\DeclareMathAccent{\widecheck}{0}{mathx}{"71}
\DeclareMathAccent{\wideparen}{0}{mathx}{"75}

\newtoggle{long}

\theoremstyle{definition}
\newtheorem{example}{Example}
\AtEndEnvironment{example}{\null\hfill\qedsymbol}%
\theoremstyle{plain}
\newtheorem{theorem}{Theorem}
\newtheorem{corollary}{Corollary}

\newtheorem{definition}{Definition}

\algrenewcommand\alglinenumber[1]{\smaller #1:}
\makeatletter
\lst@InstallKeywords k{attributes}{attributestyle}\slshape{attributestyle}{}ld
\makeatother

\usepackage[scaled=0.80]{DejaVuSansMono}
\lstdefinestyle{customc}{
  belowcaptionskip=1\baselineskip,
  breaklines=true,
  xleftmargin=\parindent,
  language=C,
  showstringspaces=false,
  basicstyle=\small\ttfamily,
  keywordstyle=\small\bfseries\ttfamily\color{NavyBlue},
  commentstyle=\itshape\color{black},
  identifierstyle=\ttfamily\color{black},
  stringstyle=\itshape\color{NavyBlue},
  keywords={ map, flatMap, reduce, return, define, gaussian, gauss, step, elif, then, skip, if, else, reduceByKey, sensitiveAttribute, fairnessTarget, qualified, def},
moreattributes={ let, where}, 
attributestyle = \itshape\ttfamily\color{attributecolor}
}

\setlist[itemize]{
    topsep=.5ex,
    itemsep=0ex,
    leftmargin=1em,
}

\setlist[description]{
    labelindent=.4cm,
    style=unboxed,
    leftmargin=.4cm,
    font=\itshape,
    topsep=.5ex,
    itemsep=0ex
}

\renewcommand{\paragraph}[1]{\vspace{.03in}\noindent\textbf{\emph{#1}}~}

\everypar{\looseness=-1}

\definecolor{lightgray}{gray}{0.9}
\definecolor{midgray}{gray}{0.65}
\definecolor{darkgray}{gray}{0.3}

\newcommand{\abr}[1]{\textsc{\MakeLowercase{#1}}}

\newcommand{\abrs}[1]{\abr{#1}{\footnotesize{s}}\xspace}

\renewcommand{\vec}[1]{\boldsymbol{#1}}

\renewcommand{\leq}{\leqslant}
\renewcommand{\geq}{\geqslant}

\renewcommand{\qedsymbol}{{\footnotesize{$\blacksquare$}}}

\newcommand{\rone}{(\emph{i})~}
\newcommand{\rtwo}{(\emph{ii})~}
\newcommand{\rthree}{(\emph{iii})~}
\newcommand{\rfour}{(\emph{iv})~}
\newcommand{\rfive}{(\emph{v})~}

\definecolor{keywordcolor}{gray}{0.0}
\definecolor{attributecolor}{gray}{0.0}
\definecolor{wildcolor}{gray}{0.8}

\usepackage{relsize}

\newcommand{\true}{\emph{true}}
\newcommand{\false}{\emph{false}}

\newcommand\lin{\lstinline[mathescape,basicstyle=\ttfamily]}
\newcommand{\linn}[1] {\begin{center}\lin{1}\end{center}}
\newcommand{\sem}[1]{\llbracket #1\rrbracket}

\newcommand{\smt}{\mathcal{L}_\mathcal{T}}

\newcommand{\dist}{D}
\newcommand{\pdf}{p}
\newcommand{\pdfs}{\mathcal{D}}

\newcommand{\apprs}{\mathcal{A}}
\newcommand{\appr}{\mathit{step}}
\newcommand{\sappr}{\appr^\phi}
\newcommand{\cdf}{F}

\newcommand{\pr}[1]{\text{Pr}[#1]}

\newcommand\cond[1]{\color{MidnightBlue}{ \left\{#1 \right\} } }
\newcommand\bcond[1]{\color{MidnightBlue}{\left\{ #1 \right\}}}

\newcommand{\prog}{\mathcal{P}}
\newcommand{\pre}{\mathcal{P}_\emph{pre}}
\newcommand{\dec}{\mathcal{P}_\emph{dec}}
\newcommand{\post}{\mathit{post}}

\newcommand{\vars}{\mathcal{X}}
\newcommand{\inputs}{\vec{v}_i}
\newcommand{\outputs}{\vec{v}_o}

\newcommand{\aexp}{E}
\newcommand{\sif}{\texttt{\textbf{if}}\xspace}
\newcommand{\sthen}{\texttt{\textbf{then}}\xspace}
\newcommand{\selse}{\texttt{\textbf{else}}\xspace}
\newcommand{\gauss}{\texttt{gauss}\xspace}
\newcommand{\skips}{\ensuremath{\varepsilon}}

\newcommand{\val}{\emph{s}}

\newcommand{\assign}{\abr{ASSIGN}\xspace}
\newcommand{\passign}{\abr{PASSIGN}\xspace}
\newcommand{\tcond}{\abr{TCOND}\xspace}
\newcommand{\fcond}{\abr{FCOND}\xspace}
\newcommand{\seq}{\abr{SEQ}\xspace}

\newcommand{\eassign}{\abr{VC-ASN}\xspace}
\newcommand{\epassign}{\abr{VC-PASN}\xspace}
\newcommand{\econd}{\abr{VC-COND}\xspace}
\newcommand{\eseq}{\abr{VC-SEQ}\xspace}

\newcommand{\fs}{FairSquare\xspace}
\newcommand{\wvc}{\abr{wvc}\xspace}
\newcommand{\vol}{\textsc{vol}\xspace}
\newcommand{\volalg}{\abr{SYMVOL}\xspace}
\newcommand{\volopt}{\abr{ADF-SYMVOL}\xspace}
\newcommand{\volume}{\mathit{vol}}
\newcommand{\pair}[1]{\langle #1 \rangle}

\newcommand{\pexp}{\emph{PExp}}
\newcommand{\prob}{\emph{Prob}}

\newcommand{\verify}{\abr{VERIFY}\xspace}

\newcommand{\semprog}{\varphi_\prog}
\newcommand{\enc}{\abr{PVC}\xspace}

\newcommand{\rect}{\mbox{\mancube}}

\newcommand{\hdecomp}{\abr{HDECOMP}\xspace}
\newcommand{\sample}{\abr{HSAMPLE}\xspace}
\newcommand{\decay}{\abr{DECAY}\xspace}

\newcommand{\block}{\mathit{block}}

\pgfmathdeclarefunction{gauss}{2}{%
  \pgfmathparse{1/(#2*sqrt(2*pi))*exp(-((x-#1)^2)/(2*#2^2))}%
}

\let\originalleft\left
\let\originalright\right
\renewcommand{\left}{\mathopen{}\mathclose\bgroup\originalleft}
\renewcommand{\right}{\aftergroup\egroup\originalright}

\toggletrue{long}

\begin{document}

\title{Quantifying Program Bias
}

\authorinfo{
Aws Albarghouthi, Loris D'Antoni, and Samuel Drews
}
           {
University of Wisconsin--Madison}
           {
           }

\authorinfo{ Aditya Nori
}
           {

Microsoft Research}
           {
           }

\maketitle
\begin{abstract}
  With the range and sensitivity of algorithmic decisions
  expanding at a break-neck speed,
  it is imperative that we aggressively investigate
  whether programs are \emph{biased}.
  We propose a
  novel probabilistic program analysis technique
  and apply it to quantifying bias in decision-making programs.
  Specifically, we \rone present a sound and complete
  automated verification technique for proving quantitative
  properties of probabilistic programs;
  \rtwo show that certain notions of bias,
  recently proposed in the \emph{fairness} literature,
  can be phrased as quantitative correctness properties; and
  \rthree present \fs, the first
  verification tool
  for quantifying program bias, and evaluate it on
  a range of
  decision-making
  programs.
%
%
\end{abstract}

\section{Introduction}\label{sec:intro}

A number of very interesting applications of program analysis
have been explored in the probabilistic setting:
reasoning about cyber-physical systems~\cite{sankaranarayanan13},
proving differential privacy of complex algorithms~\cite{barthe14},
reasoning about approximate programs and hardware~\cite{sampson2014expressing,carbin13},
synthesizing control programs~\cite{chaudhuri14},
amongst many others.
In this paper, we turn our attention to the problem
of \emph{quantifying program bias}.

\paragraph{Program bias}
Programs that make decisions can be biased.
Consider, for instance,
automatic grading of
writing prompts for standardized tests~\cite{attali06};
some speech patterns may be characterized
as poor writing style and result in lower scores.
However, if such speech patterns are affiliated with
a specific ethnic group, then the bias is a
potential source of concern.

Programs have become powerful arbitrators
of a range of significant decisions with far-reaching
societal impact---hiring~\cite{articleTimesHiring,articleWired},
welfare allocation~\cite{articleSlateWelfare},
prison sentencing~\cite{articlePropublica},
policing~\cite{articleGuardianCrime,perry2013predictive},
amongst many others.
With the range and sensitivity of algorithmic decisions expanding by the day,
the problem of understanding the nature of program bias is a pressing
one:
Indeed,
the notion of \emph{algorithmic fairness}
has recently captured
the attention of a broad spectrum of
experts, within computer science and without~\cite{dwork12,zemel13,feldman15,calders10,datta2015automated,articlePropublica,articWsjstaples,sweeney2013discrimination,tutt2016fda,ajunwa2016hiring,barocas2014big}.

Fairness and justice
have always been a ripe topic for philosophical debate~\cite{rawls2009theory},
and, of course, there are no established rigorous definitions.
Nonetheless, the rise of automated decision-making
prompted the introduction of
a number of formal definitions of fairness,
and their utility within different contexts is being actively
debated~\cite{feldman15,ruggieri2014using,dwork12,hardt16,friedler16}.
Notable formulations of fairness include
\emph{individual fairness}, which dictates
that \emph{similar} inputs must result in \emph{similar} outputs,
and \emph{group fairness}, which dictates
that a particular subset of inputs must have a similar aggregate output
to the whole.
In this paper,
we view such notions of fairness
 as \emph{quantitative properties} of decision-making
 programs.


\paragraph{Bias as a probabilistic property}
We think of decision-making algorithms as probabilistic programs, in the sense
that they are invoked on
input drawn from a probability distribution, e.g.,
representing the demographics of some population.
Quantifying program bias becomes a matter of reasoning about
probabilities of program executions.

Consider a hiring program $\prog$ that takes as input a vector
of arguments $\vec{v}$ representing a job applicant's record.
One of the arguments $v_s$ in the vector $\vec{v}$ states whether the person
is a member of a protected minority or not,
and similarly $v_q$ in $\vec{v}$ states whether
the person is qualified or not.
Evaluating $\prog(\vec{v})$ returns a Boolean value indicating
whether a person is hired.
Our goal may be to prove a group fairness property that is augmented
with a notion of qualification---that
the algorithm is \emph{just as likely} to hire a qualified
minority applicant as it is for other qualified non-minority applicants.
Formally, we state this probabilistic condition as follows:
\[
\frac{\pr{\prog(\vec{v}) = \true \mid v_s = \true \land v_q = \true}}
    {\pr{\prog(\vec{v}) = \true \mid v_s = \false \land v_q = \true}}
>
1 - \epsilon
\]
Here, $\epsilon$ is a small prespecified value.
In other words, the probability of hiring
a person $\vec{v}$, conditioned on them being a qualified minority,
is very close to (or greater than) the probability of hiring a person conditioned
on them of being a qualified non-minority.
The goal of this paper is to propose techniques for automatically proving whether a program satisfies
this kind of probabilistic properties.


Proving statements of the above form amounts
to quantifying probabilities of return outcomes of the program.
We propose an automated verification technique
that reduces the verification problem
to that of computing the \emph{weighted volume} of
the logical encoding of a program in real arithmetic.
We then utilize a novel \emph{symbolic volume computation} algorithm
that exploits the power of \abr{SMT} solvers
to integrate probability density functions over
real regions defined by program encodings.
We show that our algorithm is
guaranteed to converge to the exact values in the limit,
thus resulting in a sound and complete verification procedure.
To our knowledge, this is one of the first probabilistic inference algorithms for \abr{SMT} with this expressivity and guarantees.
We implement our algorithm in a tool called \fs, which we
evaluate on a number
  decision-making
  programs generated by a range of machine-learning
  algorithms from real-world data.

\paragraph{Contributions}
This paper makes a number of conceptual, algorithmic, and practical contributions:
\begin{itemize}
  \item
  We present a  verification
  technique for probabilistic programs
  that reduces the problem to a set of
  weighted volume computation problems.
   We present a novel weighted-volume-computation algorithm,
   for formulas over \emph{real closed fields},
  that utilizes an \abr{SMT} solver as a black box, and
  we prove that it
  converges to the exact volume in the limit.
  To our knowledge, this is one of the first  probabilistic
  inference algorithms for \abr{SMT} with this generality and guarantees.
  (Sec.~\ref{sec:vol})

  \item We present an automated verification tool, \fs, and
  use it to quantify certain types of bias in a broad spectrum
  of programs representing machine-learning classifiers
  generated from real-world datasets.
  Our evaluation demonstrates the power of our technique
  and its ability to outperform state-of-the-art
  probabilistic program analyses.
  (Sec.~\ref{sec:impl})

\end{itemize}

\iftoggle{long}{}{
The \emph{supplementary materials} include a full version
of this paper with proofs, as well as detailed experimental results
and the benchmark programs from Section~\ref{sec:impl}.
}


\section{Overview and Illustration}\label{sec:example}

Our problem setting is as follows:
First, we are given a \emph{decision-making program} $\prog_\emph{dec}$.
Second, we have a \emph{probabilistic precondition} defining a probability
distribution over inputs of $\prog_\emph{dec}$.
We define the probability distribution operationally as a probabilistic
program $\prog_\emph{pre}$, which we call the \emph{population model}.
Intuitively, the population model provides a probabilistic picture
of the population from which the inputs of  $\prog_\emph{dec}$ are drawn.
Third, we are given a quantitative postcondition $\post$ that correlates the probabilities of various program outcomes.
This postcondition can encode various properties
relating to program bias;
intuitively, our goal is to prove the following triple:
$${\cond{\vec{v} \sim \prog_\emph{pre}}}\quad
r \gets \prog_\emph{dec}(\vec{v})\quad
{\bcond{\post
}}$$

Let us consider various possible instantiations of $\post$.
Feldman et al.~\cite{feldman15} introduced the following
definition, inspired by \emph{Equality of Employment
Opportunity Commission's}~\cite{EEOC} recommendation
in the US:
\[
   \frac{\pr{r = \true  \mid \emph{min}(\vec{v}) = \true}}
    {\pr{r = \true  \mid \emph{min}(\vec{v}) = \false}}
>
1 - \epsilon
\]
Assuming $\prog_\emph{dec}$ returns a Boolean value---indicating
whether an applicant $\vec{v}$ is hired---%
this \emph{group fairness} property states that the selection rate
from a minority group, $\emph{min}(\vec{v}) = \true$, is as good as the selection rate from the
rest of the population.
One can thus view this verification problem
as proving a probabilistic property involving
\emph{two sets of program traces}: one set where the input $\emph{min}(\vec{v})$ is true,
and another where it is false.
Alternatively, the above definition could be strengthened
with a lower bound on the ratio, so as
to ensure that the selection rate of the two groups is similar (\emph{statistical parity}).
Further, we could additionally condition on \emph{qualified}
applicants, e.g., if the job has some minimum qualification,
we do not want to characterize group fairness for arbitrary
applicants, but only within the qualified subpopulation.
Various comparable notions of group fairness
have been proposed and used in the literature, e.g., ~\cite{feldman15,zemel13,datta2016algorithmic}.

While the above definition is concerned with fairness at the level
of subsets of the domain of the decision-making program,
\emph{individual fairness}~\cite{dwork12} is concerned with similar outcomes
for similar elements of the domain.
In our hiring example, one potential formulation is as follows:
\[
\pr{r_1 = r_2  \mid \vec{v}_1 \sim \vec{v}_2 } > 1 - \epsilon
\]
In other words, we want to ensure that for any two
individuals, if they are \emph{similar} ($\sim$),
then we want them to receive similar outcomes ($r_1 = r_2$) with a high probability.
This is a \emph{hyperproperty}---as it considers two copies
of $\prog_\emph{dec}$---and can be encoded
through \emph{self-composition}~\cite{barthe04}.
This property is close in nature to differential privacy~\cite{dwork2006differential}
and robustness~\cite{Chaudhuri11,bastani16}.

Of course, various definitions of fairness have
their merits and their shortcomings,
and there is an ongoing discussion on this subject~\cite{friedler16,dwork12,hardt16,feldman15,ajunwa2016hiring}.
Our contribution is not to add to this debate,
but to cast fairness as a quantitative property of programs, and
therefore enable automated reasoning about fairness
of decision-making programs.

\begin{figure}[t]
  \centering
\fbox{\includegraphics[scale=1.0]{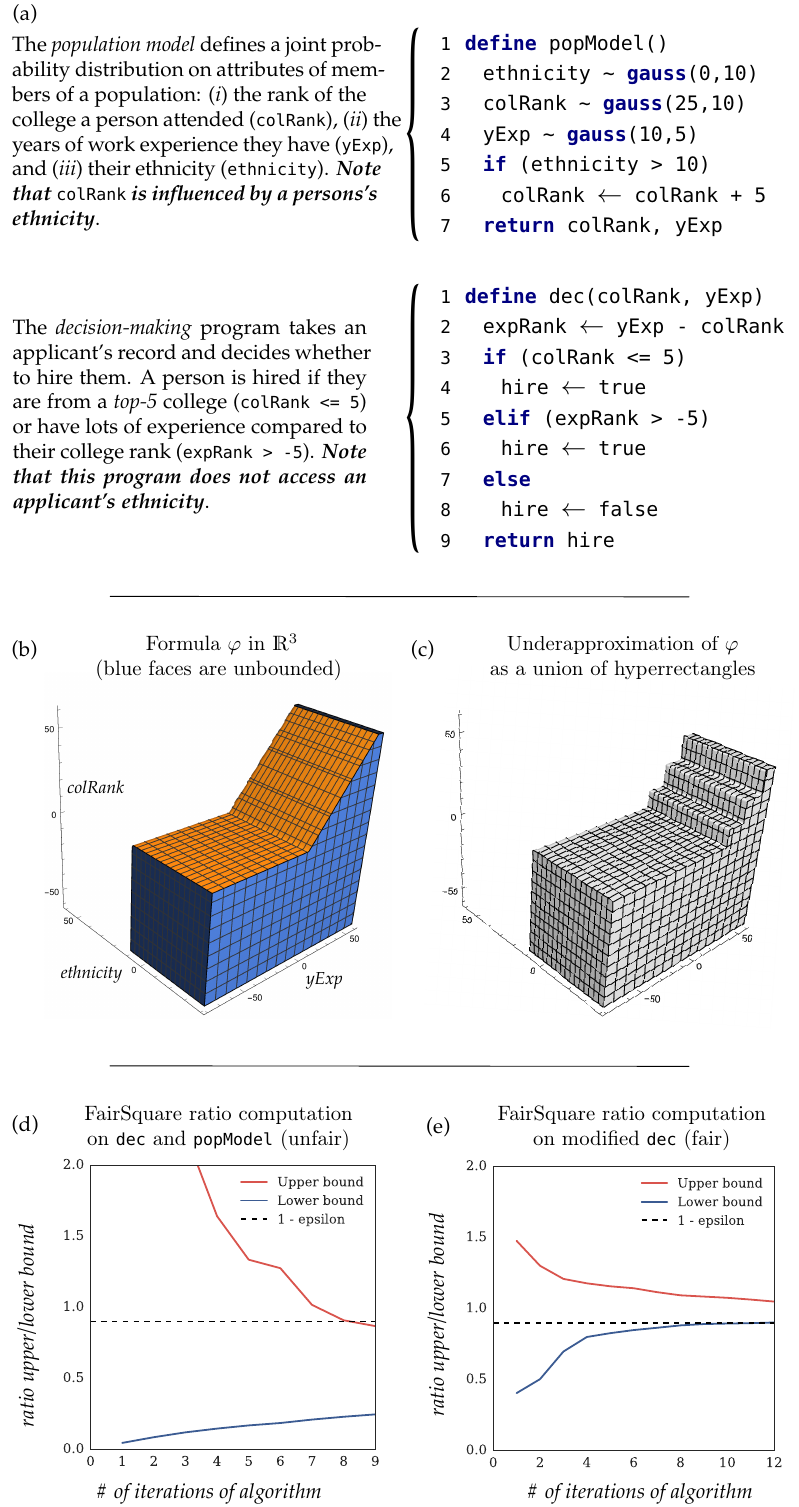}}
\caption{Simple illustrative example}
\label{fig:ex}
\end{figure}

\paragraph{A simple verification problem}
Consider the two programs in Figure~\ref{fig:ex}(a).
The program \texttt{popModel} is a probabilistic
program describing a simple model of the population.
%
Here, a member of the population has three
attributes, all of which are real-valued:
\rone \texttt{ethnicity};
\rtwo \texttt{colRank}, the ranking of the college
the person attended (lower is better);
and \rthree \texttt{yExp}, the years of work experience
a person has.
We consider a person is a member of a protected group
if \texttt{ethnicity > 10}; we call this the
\emph{sensitive condition}.
The population model can be viewed as a \emph{generative model}
of records of individuals---the more likely a combination
is to occur in the population, the more likely it will be generated.
For instance, the years of experience an individual has (line 4)
follows a Gaussian (normal)
distribution with mean $10$ and standard deviation $5$.
Observe that our model specifies that
members of a protected minority will probably
attend a lower-ranked college, as encoded in lines 5-6.

The program \texttt{dec} is a decision-making
program that takes a job applicant's
college ranking and years of experience and
decides whether they get hired (the \emph{fairness target}).
The program implements a decision tree,
perhaps one generated by a machine-learning algorithm.
A person is hired if they attended a \emph{top-5}  college (\texttt{colRank <= 5}) or  have lots of experience compared to their college's ranking (\texttt{expRank > -5}).
Observe that \texttt{dec} \emph{does not access an applicant's ethnicity}.

Our goal is to establish
whether
the hiring algorithm \texttt{dec}  discriminates against
members of the protected minority.
Concretely, we attempt to prove the following property:
\[
   \frac{\pr{\texttt{hire}  \mid \texttt{min}}}
    {\pr{\texttt{hire} \mid \neg \texttt{min}}}
>
1 - \epsilon
\]
where \texttt{min} is shorthand for the sensitive condition
\texttt{ethnicity > 10},
and  $\epsilon$ is a small parameter set to $0.1$ for illustration.
%
Despite the potential shortcomings of this group fairness property~\cite{dwork12},
its simple formulation serves well as an illustration of our technique.

We can rewrite the above statement to eliminate
conditional probabilities as follows:
\begin{align}\label{al:fair}
\frac{\pr{\texttt{hire} \land \texttt{min}} \cdot \pr{\neg \texttt{min}}}
    {\pr{\texttt{hire} \land \neg \texttt{min}} \cdot \pr{\texttt{min}}}
>
1 - \epsilon
\end{align}
Therefore, to prove the above statement,
we need to compute a value for each of the probability
terms: $\pr{\texttt{hire} \land \texttt{min}}$, $\pr{\texttt{min}}$,
and $\pr{\texttt{hire} \land \neg \texttt{min}}$.
(Note that $\pr{\neg \texttt{min}} = 1 - \pr{\texttt{min}}$.)

Notice that, to prove or disprove inequality~\ref{al:fair},
all we need are good enough bounds on the values of such probabilities
and not their  exact values.

For the purposes of illustration, we shall focus
our description on computing $\pr{\texttt{hire} \land \neg \texttt{min}}$.

\paragraph{Probabilistic verification conditions}
To compute the probability $\pr{\texttt{hire} \land \neg \texttt{min}}$,
we need to reason about the \emph{composition} of the two
programs, $\texttt{dec}\circ\texttt{popModel}$.
That is, we want to compute the probability that
\rone \texttt{popModel} generates a non-minority applicant,
and \rtwo \texttt{dec} hires that applicant.
To do so, we begin by encoding both programs as formulas
in the linear real arithmetic theory of first-order logic.
The process is analogous to that of standard \emph{verification
condition} (\abr{VC}) generation for loop-free program fragments.

First, we encode \texttt{popModel} as follows:
\begin{align*}
\varphi_\texttt{pop} \equiv&~
\emph{ethnicity} > 10 \Rightarrow \emph{colRank}_1 = \emph{colRank} + 5 \\
&\land \emph{ethnicity} \leq 10 \Rightarrow \emph{colRank}_1 = \emph{colRank} \end{align*}
where subscripts are used to encode multiple occurrences
of the same variable (i.e., \abr{SSA} form).
Note that assignments in which values are drawn from probability distributions do not appear in the encoding---we shall
address them later.

Second, we encode \texttt{dec} as follows (after simplification):
\begin{align*}
\varphi_\texttt{dec} \equiv&~
  \emph{expRank} = \emph{yExp}^i - \emph{colRank}^i\\
  &\land
  \emph{hire} \iff (\emph{colRank}^i \leq 5 \lor \emph{expRank} > -5)
\end{align*}
where variables with the superscript $i$ are the input
arguments to $\texttt{dec}$.
Now, to encode the composition $\texttt{dec} \circ \texttt{popModel}$,
we simply conjoin the two formulas---$\varphi_\texttt{pop}$ and $\varphi_\texttt{dec}$---and add equalities
between returns of \texttt{popModel} and arguments of \texttt{dec}.
$$\semprog \equiv \varphi_\texttt{pop} \land \varphi_\texttt{dec} \land
\emph{yExp}^i = \emph{yExp} \land \emph{colRank}^i = \emph{colRank}_1$$

Our goal is to compute the probability that a
non-minority applicant gets hired.
Formally, we are asking,
\emph{what is the probability that the following formula is satisfied?}
$$\varphi \equiv \exists V_d \ldotp \semprog  \land \emph{hire} \land \emph{ethnicity}\leq10$$
 $V_d$ is the set of variables
that are not probabilistically assigned to, that is, all variables \emph{other than}
the three variables
$V_p = \{\emph{ethnicity}, \emph{colRank},\emph{yExp}\}$.
Intuitively, by \emph{projecting out} all non-probabilistic
variables, we get a formula $\varphi$ whose models are
the set of all probabilistic samplings
that lead to a non-minority applicant being generated and hired.

\paragraph{Weighted volume computation}
To compute the probability that $\varphi$
is satisfied, we begin by noting that
$\varphi$ is, geometrically,
a region in $\mathds{R}^3$, because it has three free real-valued variables, $V_p$.
The region $\varphi$ is partially illustrated in Figure~\ref{fig:ex}(b).
Informally, the probability of satisfying $\varphi$
is the probability of drawing values for the dimensions in $V_p$
that end up \emph{falling} in the region $\varphi$.
Therefore, the probability of satisfying $\varphi$
is its \emph{volume} in $\mathds{R}^3$,
\emph{weighted} by the probability density of each of the
three variables.
Formally:
$$\textstyle \pr{\texttt{hire} \land \neg \texttt{min}} = \int_{\varphi} p_ep_yp_c ~dV_p$$
where, e.g., $p_e$ is the \emph{probability density function}
of the distribution \texttt{gauss(0,10)}---the distribution
from which the value of \texttt{ethnicity} is drawn  in line 2 of \texttt{popModel}. Specifically, $p_e$ is a function of $\emph{ethnicity}$,
namely, $p_e(\emph{ethnicity}) = \frac{1}{10\sqrt{2\pi}}e^{-\frac{\emph{ethnicity}^2}{200}}$.

The primary difficulty here is that the region of integration is specified by an
arbitrary \abr{SMT} formula over an arithmetic theory.
So, \emph{how do we compute a numerical value for this integral?}
We make two interdependent observations:
\rone if the formula represents a \emph{hyperrectangular region}
in $\mathds{R}^n$---i.e., a box---then integration is typically simple,
due to the constant upper/lower bounds of all dimensions;
\rtwo we can \emph{symbolically decompose} an \abr{SMT}
formula into an (infinite) set of hyperrectangles.

Specifically, given our formula $\varphi$,
we  construct a new formula,
$\rect_\varphi$, where each model $m \models \rect_\varphi$
corresponds to a hyperrectangle that
underapproximates $\varphi$.
Therefore, by systematically finding disjoint hyperrectangles
inside of $\varphi$ and computing
their weighted volume, we iteratively improve a
\emph{lower bound} on the exact weighted volume of $\varphi$.
Figure~\ref{fig:ex}(c) shows a possible
underapproximation of $\varphi$ composed of four hyperrectangles.
Sec.~\ref{sec:vol} formalizes this technique  and proves
its convergence for decidable arithmetic theories.

%

\paragraph{Proofs of group fairness}
We demonstrated how our technique reduces the problem
of computing probabilities to weighted volume computation.
%
Figure~\ref{fig:ex}(d) illustrates a run of  our tool, \fs,  on this example.
\fs iteratively
improves lower and upper bounds for the
probabilities in the ratio,
and, therefore,
the ratio itself.
Observe how the upper bound (red) of the ratio is decreasing
and its lower bound (blue) is increasing.
This example is not group fair for $\epsilon = 0.1$,
since the upper bound goes below $0.9$.

Recall that applicants of a protected minority
tend to attend lower-ranked colleges,
as defined by \texttt{popModel}.
Looking at \texttt{dec}, we can point out that the
cause for unfairness is the importance of college ranking
for hiring.
Let us attempt to fix this by modifying line 2 of \texttt{dec}
to \lstinline{expRank $\gets$ 5*yExp - colRank}.
In other words, we have made the hiring algorithm value an applicant's job
experience way more than college ranking.
The run of \fs on the modified \texttt{dec} is illustrated
in Figure~\ref{fig:ex}(e), where the lower bound on the ratio
exceeds 0.9, thus proving our fairness property.


\section{Probabilistic Programs and Verification}\label{sec:prelims}

%
%
%
We formally define programs
and present a general framework for stating and verifying probabilistic
properties.
%
%

\subsection{Program model and  semantics}
\paragraph{Programs}
A program $\prog$ is a sequence of statements $S$:
\begin{align*}
    S \coloneqq & ~V \gets \aexp & \text{ assignment statement}\\[-3pt]
        \mid&~ V \sim \dist &\text{ probabilistic assignment}\\[-3pt]
        \mid&~ \sif~ B ~\sthen~ S ~\selse~ S & \text{conditional}\\[-3pt]
        \mid&~ SS & \text{ sequence of statements}
\end{align*}
where $V$ is the set of \emph{real-valued variables} that can appear
in $\prog$, $e \in \aexp$ is an arithmetic expression over variables in $V$,
and $b \in B$ is a Boolean expression over variables in $V$.
A probabilistic assignment is made by sampling from
a probability distribution $p \in D$. A probability distribution
can be, for example,
a \emph{Gaussian distribution}, denoted by $\gauss(\mu, \sigma)$, where
$\mu, \sigma \in \mathds{R}$ are the mean and standard deviation
of the Gaussian.
We shall  restrict distributions to be \emph{univariate} and with
constant parameters, e.g., mean and standard deviation of a Laplacian
or Gaussian---that is, we assume independence of probabilistic assignments.
Given a probabilistic assignment $x \sim p$, we shall
treat $p(x)$ as a \emph{probability density function} (\abr{PDF})
of the distribution from which the value assigned to $x$ is drawn.
For instance, if the distribution is \texttt{gauss(0,1)},
then $p(x) = \frac{1}{\sqrt{2\pi}}e^{-\frac{\emph{x}^2}{2}}$.
%

We use $\vec{v}_i$ to denote a vector of \emph{input variables}
of $\prog$, and $\vec{v}_o$ to denote a vector of \emph{output variables}
of $\prog$; these variables appear in $V$ and
denote the arguments and returns of $\prog$.
We say that a program is \emph{closed} if it has no inputs, i.e.,
$\vec{v}_i$ is empty.
We shall  refer to the following subsets of $V$.
\begin{itemize}
    \item $V_p \subseteq V$ is the set of \emph{probabilistic variables}:
    those that get assigned to in probabilistic assignments.
    \item $V_d = V \setminus V_p$ is the set of \emph{deterministic variables}:
    those that do not appear in probabilistic assignments.
\end{itemize}

This simple language
can be used to describe decision-making programs
 typical machine-learning classifiers
such as decision trees, support vector machines, Bayesian networks, neural networks,
as well as loop-free probabilistic programs (loops with constant bounds
can be unrolled).%
\footnote{Unbounded loops
could be handled through iterative unrolling
as in~\cite{sankaranarayanan13}.}
As demonstrated in Sec.~\ref{sec:example},
the same language is used to define probabilistic preconditions programmatically.

\paragraph{Operational semantics}
Typically,
the \emph{state} $\val : V \rightarrow \mathds{R}$  of the program is defined
as a valuation function from variables in $V$ to
values in $\mathds{R}$.
In a probabilistic setting, however, we need to maintain
an additional state that dictates values drawn from probability distributions.
Following standard semantics of probabilistic programs~\cite{kozen1981semantics},
we assume a finite sequence of independent
 random variables.
The semantics of an execution is thus defined for a
fixed sequence of values $\omega$ of these variables.
Informally, it is as if we performed all sampling
before the program executes and stored the results
in a sequence for use whenever we encounter a probabilistic
assignment. For the full  semantics, refer to
\iftoggle{long}{Appendix~\ref{app:opsem}}{\emph{supplementary materials}}.

%

%

\subsection{Programs and volume computation}

Following Chistikov et al.~\cite{chistikov14arxiv}, we reduce the
problem of computing the probability that the program
terminates in a state satisfying $\varphi$
to \emph{weighted volume computation} (\wvc) over formulas
 describing regions in $\mathds{R}^n$.
In what follows, we begin by formalizing the \wvc problem.

\paragraph{Volume of a formula}
We will use $\smt$ to denote  first-order
formulas in some real arithmetic theory $\mathcal{T}$.
Specifically, we consider two decidable theories:
\emph{linear real arithmetic} and the strictly
richer \emph{real closed fields}---Boolean combinations
of polynomial inequalities.
Given a formula $\varphi \in \smt$, a model $m$ of $\varphi$,
denoted by $m \models \varphi$, is a point in $\mathds{R}^n$,
where $n$ is the number of free variables of $\varphi$.
Thus, we view $\varphi$ as a region in $\mathds{R}^n$, i.e.,
$\varphi \subseteq \mathds{R}^n$.
We use $\vars_\varphi = \{x_1,\ldots,x_n\}$
to denote the free variables
of $\varphi$.

The (unweighted) \emph{volume} of a formula $\varphi$ is
$\int_\varphi 1~ d\vars_\varphi$
where $d\vars_\varphi$ is short for $dx_1dx_2\ldots dx_n$.
For example, if $\varphi$ is in $\mathds{R}^2$, then
$\int_\varphi 1~d\vars_\varphi$ is the area of $\varphi$.

\paragraph{Weighted volume of a formula}
We now define the \emph{weighted volume} of a formula.
We assume we are given a pair $(\varphi, \pdfs)$,
where $\varphi \in \smt$ and $\pdfs = \{\pdf_1,\ldots,\pdf_n\}$ is a set
of  probability density functions
such that each variable $x_i \in \vars_\varphi$ is
associated with a density function $\pdf_i(x_i)$
of the probability distribution of its values.
The weighted volume of $\varphi$ with respect to $\pdfs$, denoted by $\vol(\varphi,\pdfs)$,
is defined as follows:
$$\textstyle \int_\varphi \prod_{x_i \in \vars_\varphi} \pdf_i(x_i) ~d\vars_\varphi$$
%
\begin{example}
  Consider the formula $\varphi \equiv x_1 + x_2 \geq 0$,
  and let $\pdfs = \{p_1,p_2\}$, where $p_1$ and $p_2$ are
  the \abr{PDF} of the
  Gaussian distribution with mean $0$ and standard deviation $1$.
  Then,
  $$\textstyle \vol(\varphi,\pdfs) = \int_{x_1+x_2\geq0} p_1(x_1)p_2(x_2)~dx_1dx_2 = 0.5$$
  Intuitively, if we are to randomly draw two values for $x_1$ and $x_2$
  from the Gaussian distribution,
  we will land in the region $x_1+x_2\geq 0$ with  probability 0.5.
\end{example}

\paragraph{Probabilistic verification conditions}
Recall that our goal is to compute the probability
of some predicate $\varphi$ at the end of a program
execution, denoted $\pr{\varphi}$.
We now show how to encode this problem
as weighted volume computation.
First, we encode program executions
as a formula $\semprog$.
The process is similar to standard
verification condition generation (as used by verification~\cite{barnett2005weakest}
and bounded model checking tools~\cite{clarke2004tool}), with the difference
that probabilistic assignments populate a set $\pdfs$
of probability density functions.

\begin{figure}[t]
  \centering
  \smaller
  \begin{prooftree}
  \justifies
  \pair{x = \sem{e}, \emptyset} \rhd x \gets e
  \using \eassign
  \end{prooftree}
%
  \hspace{1em}
  \begin{prooftree}
  \justifies
  \pair{\true, \{\pdf_i\}} \rhd x_i \sim \pdf_i
  \using \epassign
  \end{prooftree}

  \vspace{.1in}
  \begin{prooftree}
  \pair{\varphi_1,\pdfs_1} \rhd S_1
  \quad\quad
  \pair{\varphi_2,\pdfs_2} \rhd S_2
  \justifies
  \pair{\varphi_1 \land \varphi_2, \pdfs_1 \cup \pdfs_2} \rhd S_1S_2
  \using \eseq
  \end{prooftree}

  \vspace{.1in}
  \begin{prooftree}
  \pair{\varphi_1,\pdfs_1} \rhd S_1
  \quad\quad
  \pair{\varphi_2,\pdfs_2} \rhd S_2
  \justifies
  \pair{\emph{ite}(\sem{b},\varphi_1,\varphi_2),\pdfs_1 \cup \pdfs_2} \rhd \sif~b~\sthen~S_1~\selse~S_2
  \using \econd
  \end{prooftree}
  \caption{Probabilistic verification condition generation.
  Above, $\emph{ite}(a,b,c) \triangleq (a \Rightarrow b) \land (\neg a \Rightarrow c)$.}
  \label{fig:enc}
\end{figure}

 Figure~\ref{fig:enc} inductively defines
 the construction of a \emph{probabilistic verification condition} for a program $\prog$, denoted by a function $\enc(\prog)$, which returns a pair $\pair{\semprog, \pdfs}$.
Without loss of generality, to simplify our exposition,
we assume programs
are in \emph{static single assignment} (\abr{SSA}) form~\cite{cytron1991efficiently}. Given a Boolean expression $b$,
the denotation $\sem{b}$ is the same expression
interpreted as an $\smt$ formula.
The same applies to arithmetic expressions $e$.
For example, $\sem{\texttt{x + y > 0}} \triangleq x + y > 0$.
Intuitively, the construction generates a formula $\varphi_\prog$
that encodes program executions, treating probabilistic assignments
as non-deterministic,
and a set $\pdfs$ of the \abrs{PDF} of distributions in probabilistic
assignments (rule \epassign).

Now, suppose we are given a closed program $\prog$ and a
Boolean formula $\varphi$ over its output variables. Then,
$$\pr{\varphi} = \vol(\exists V_d \ldotp \semprog \land \varphi,\pdfs)$$
That is, we project out all non-probabilistic variables
from $\semprog \land \varphi$ and compute the weighted
volume with respect to the densities $\pdf_i \in \pdfs$.
Intuitively, each model $m$ of $\exists V_d \ldotp \semprog \land \varphi$
corresponds to a sequence of values drawn
in probabilistic assignments in an execution of $\prog$.
We note that our construction is closely
related to that of Chistikov et al.~\cite{chistikov14arxiv},
to which we refer the reader for a measure-theoretic
formalization.

\begin{example}
Consider the following closed program $\prog$
\begin{lstlisting}
  x ~ gauss(0,2); y ~ gauss(-1,1); z $\gets$ x + y
\end{lstlisting}
where \texttt{z} is the return variable.
Using the encoding in Figure~\ref{fig:enc},
we compute the pair $\pair{\semprog,\pdfs} \rhd \prog$, where
$\semprog \triangleq z = x + y$
and
$\pdfs = \{\pdf_x, \pdf_y\}$,
where $\pdf_x$ and $\pdf_y$ are the \abrs{PDF} of the
two distributions from which values of $x$ and $y$
are drawn.

Suppose that we would like to compute the probability
that \texttt{z} is positive when the program terminates:
$\pr{z \geq 0}$.
Then, we can compute the following weighted volume:
$\vol(\exists z \ldotp \semprog \land z \geq 0, \pdfs)$,
which is equal to $0.32736$.
\end{example}

\subsection{Probabilistic verification problems}
We now define probabilistic verification problems and present
an abstract verification algorithm that assumes the existence of an oracle
for weighted volume computation.

\paragraph{Verification problems}
A verification problem is a triple
$(\pre, \dec, \post)$,
where
\begin{itemize}
  \item  $\pre$,  called the \emph{probabilistic precondition},
   is a closed program
  over variables $V^\emph{pre}$ and
  output variables $\outputs^\emph{pre}$.

  \item $\dec$, called the \emph{decision-making program},
  is an open program over variables $V^\emph{dec}$;
  its input arguments are $\inputs^\emph{dec}$,
  with $|\inputs^\emph{dec}| = |\outputs^\emph{pre}|$;
  and its output variables are $\outputs^\emph{dec}$.
  (We assume that $V^\emph{pre}\cap V^\emph{dec}=\emptyset$.)

  \item $\post$ is a \emph{probabilistic postcondition},
  which is a Boolean expression over probabilities of program outcomes.
  Specifically, $\post$ is defined as follows:
  \newcommand{\oper}{X}
  \begin{align*}
    \post \in \pexp \coloneqq&~ \prob > \prob \mid \pexp \lor \pexp\\
          \mid&~ \pexp \land \pexp \mid \neg \pexp\\
    \prob \coloneqq&~ \pr{\varphi} \mid \pr{\varphi \mid \varphi} \mid c
          \mid \prob \odot \prob\\
    \odot \in&~ \{+,-,\div,\times\} ~~~~ c \in \mathds{R}
  \end{align*}
  where $\varphi \in \smt$ are formulas over input and output
  variables of $\dec$.
  For example, \emph{post} might be of the form
  $$\pr{x > 0} > 0.5 \land \pr{y + z > 7} > \pr{t > 5}$$
\end{itemize}
The goal of verification is to prove that $\post$ is true
for the  program $\dec \circ \pre$, i.e.,
the composition of the two programs where we first
run $\pre$ to generate an input for $\dec$.
Since $\pre$ is closed, the program $\dec \circ \pre$ is also closed.


\paragraph{Verification algorithm}
We now describe an idealized verification
algorithm that assumes the existence of an oracle
for weighted volume computation.
The algorithm, \verify, shown in Figure~\ref{alg:verify},
takes a verification problem and returns
whether the probabilistic postcondition holds.

\verify begins by encoding the composition of the two programs, $\dec \circ \pre$, as the pair $\pair{\varphi_\prog, \pdfs}$
and adds the constraint
$\inputs^\emph{dec} = \outputs^\emph{pre}$
to connect the outputs of $\pre$ to the inputs of $\dec$
(recall the example from Sec.~\ref{sec:example} for an illustration).
For each term of the form $\pr{\varphi}$ appearing in $\post$,
the algorithm computes its numerical value and maintains it in a map $m$.
If $m$ satisfies the $\post$---i.e., by replacing all
terms $\pr{\varphi}$ with their values in $m$---then the
postcondition holds.

\begin{figure}[!t]
  \small
  \begin{spacing}{1}
    \begin{algorithmic}[1]
        \Function{\verify}{$\pre, \dec, \post$}
        \State $\pair{\varphi_\emph{pre},\pdfs_\emph{pre}} \gets \enc(\pre)$
        \State $\pair{\varphi_\emph{dec},\pdfs_\emph{dec}} \gets \enc(\dec)$
        \State $\pair{\varphi_\prog,\pdfs} \gets
          \pair{\varphi_\emph{pre}\land \varphi_\emph{dec} \land \inputs^\emph{dec} = \outputs^\emph{pre}, \pdfs_\emph{pre} \cup \pdfs_\emph{dec}}$
        \State $V_d \gets V_d^\emph{pre} \cup V_d^\emph{dec}$
        \State $m \gets \emptyset$
        \For{$\pr{\varphi} \in \post$}
          \State $m \gets m[\pr{\varphi} \mapsto \vol(\exists V_d \ldotp \semprog \land \varphi, \pdfs)]$
        \EndFor
        \vspace{-3pt} 
        \State \Return $m \models \post$
        \EndFunction
    \end{algorithmic}
    \caption{Abstract verification algorithm}\label{alg:verify}
  \end{spacing}
\end{figure}


\section{Symbolic Weighted Volume Computation}\label{sec:vol}

In this section, we describe our weighted volume computation
algorithm. Recall that, given a formula  $\varphi$
and a set $\pdfs$ defining the \abrs{PDF}
of the distributions of free variables,
our goal is to evaluate the
integral
$\textstyle \int_\varphi \prod_{x_i\in \vars_\varphi}
\pdf_i(x_i) ~d\vars_\varphi$.
%

\paragraph{Existing techniques}
In general, there is no systematic technique
for computing an exact value for such an integral.
Moreover, even simpler linear versions of the volume computation problem, not involving
probability distributions, are \#P-hard~\cite{dyer1988complexity}.
Existing techniques suffer from
one or more of the following: they
\rone  restrict $\varphi$
to a conjunction of linear inequalities~\cite{sankaranarayanan13,de2012software},
\rtwo restrict integrands to polynomials or simple distributions~\cite{de2012software,belle15,chistikov15,belle2015hashing},
\rthree compute approximate solutions with probabilistic guarantees~\cite{chistikov14arxiv,chistikov15,vempala2005geometric,belle2015hashing},
\rfour restrict $\varphi$ to bounded regions of $\mathds{R}^n$~\cite{chistikov14arxiv,chistikov15},
or \rfive have no convergence guarantees, e.g., computer algebra
tools that find closed-form solutions~\cite{matlab,mathematica,Gehr16}.
(See Sec.~\ref{sec:relwork} for details.)

\paragraph{Symbolic weighted volume computation}
Our approach is novel in its generality and its algorithmic core.
The following are the high-level properties of our algorithm:
\begin{enumerate}
  \item Our approach accepts formulas in the decidable yet rich
 theory of \emph{real closed fields}:
  Boolean combinations of polynomial inequalities.

  \item Our approach imposes no restrictions on the form of the
  \abrs{PDF}, only
  that we can evaluate the \emph{cumulative distribution
  functions}\footnote{The \emph{cumulative distribution function} of a real-valued random variable $X$ is the function
  $f:\mathds{R}\rightarrow\mathds{R}$, such that $f(x)=\pr{X\leq x}$.}
  (\abrs{CDF}) associated with the \abrs{PDF}
  in $\pdfs$.

  \item Our approach is guaranteed to converge to the exact
  value of the weighted volume in the limit,
  allowing us to produce a sound and complete verification procedure.
\end{enumerate}

At the algorithmic level, our approach makes the following contributions:
\begin{enumerate}
  \item Our approach exploits the power of \abr{SMT}
  solvers and uses them as a black box,
  allowing it to directly benefit from future advances in solver technology.
  \item Our approach employs the idea of dividing the space
  into rectangular regions that are easy to integrate over.
  While this age-old idea has been employed in various guises in
  verification~\cite{Bournez99, sankaranarayanan13,asarin2000approximate,li2014management},
  we utilize it in a new symbolic way to enable volume computation over
  \abr{SMT} formulas.
  \item Our approach introduces a novel technique for approximately encoding
  \abrs{PDF} as formulas, and using them to guide the \abr{SMT}
  solver towards making large leaps to the exact solution.
\end{enumerate}


\begin{figure}[t]
  \small
    \centering
\begin{prooftree}
\justifies
\volume \gets 0
\quad \Psi \gets \rect_\varphi
\using \hdecomp
\end{prooftree}

\vspace{.2in}
\begin{prooftree}
    m \models \Psi
    \justifies
    \volume \gets \volume + \vol(H^m,\pdfs)
    \quad \Psi \gets \Psi \land \block(H^m)
    \using \sample
\end{prooftree}

\vspace{.1in}

$$\text{where } \block(H^m) \equiv \bigvee_{x \in \vars_\varphi}
  u_x < H^m_l(x) \lor l_x > H^m_u(x)$$
  \caption{\volalg:  weighted volume computation algorithm}
\label{alg:vol}
\end{figure}

\subsection{Weighted Volume Computation Algorithm}
\label{sec:swv}
To compute the integral over the region
$\varphi$, we exploit the observation
that if $\varphi$ is a \emph{hyperrectangular region},
i.e., an $n$-dimensional rectangle in $\mathds{R}^n$, then we can evaluate the
integral, because each dimension has constant lower and upper bounds.
For instance, consider the following formula representing a rectangle in $\mathds{R}^2$:
$$\varphi \equiv 0 \leq x_1 \leq 100 \land 4 \leq x_2 \leq 10$$
\begin{flalign*}
    \text{The following holds:} ~~&\textstyle \int_\varphi \pdf_1(x_1)\pdf_2(x_2)~ dx_1dx_2 &\\
    =&\textstyle(\int_0^{100} \pdf_1(x_1)~ dx_1) (\int_4^{10} \pdf_2(x_2)~ dx_2) &\\
    =&\textstyle(\cdf_1(10) - \cdf_1(4))(\cdf_2(100) - \cdf_2(0))&
\end{flalign*}
where $\cdf_i$ is the \abr{CDF} of $\pdf_i(x_i)$,
which is $F_i(x) = \int^{x}_{-\infty} \pdf_i(t) ~dt$.
That is, we independently compute the integral along
each dimension of the rectangle, and take the product.
This holds since we assume all variables are independently
sampled.

Our algorithm is primarily composed of two steps:
  First, the \emph{hyperrectangular decomposition} phase
  represents the formula $\varphi$ as a set of hyperrectangles.
  Note that this set is likely to be infinite.
  Thus, we present a technique for defining all hyperrectangles that lie in $\varphi$ symbolically as a formula $\rect_\varphi$, where each model
  of $\rect_\varphi$ corresponds to a hyperrectangle that
  lies inside the region $\varphi$.
  Second, after characterizing the set $\rect_\varphi$ of all
  hyperrectangles in $\varphi$, we can iteratively
  \emph{sample hyperrectangles} in $\varphi$,
  which can be done using an off-the-shelf \abr{SMT}
  solver to find models of $\rect_\varphi$.
  For each hyperrectangle we sample, we compute its weighted
  volume and add it to our current solution.
  Therefore, the current solution maintained
  by the algorithm is the weighted volume
  of an underapproximation of $\varphi$---that is, a lower bound on the exact
  weighted volume of $\varphi$.

\paragraph{Hyperrectangular decomposition}
We begin by defining hyperrectangles as  special formulas.
\begin{definition}[Hyperrectangles and their weighted volume]
  A formula $H \in \smt$ is a hyperrectangle
  if it can be written in the form
  $\bigwedge_{x \in \vars_H} c_x \leq x \leq c_x'$
  where $c_x,c'_x \in \mathds{R}$ are the lower
  and upper bounds of dimension $x$.
  We use $H_l(x)$ and $H_u(x)$
  to denote the lower and upper bounds of $x$ in $H$.

  The weighted volume of $H$, given a set $\pdfs$, is as follows:
  $$\textstyle \vol(H,\pdfs) = \prod_{x_i \in \vars_H} \int_{H_l(x_i)}^{H_u(x_i)} p_i(x_i) ~dx_i$$
\end{definition}

%
%

\begin{figure}[t]
  \centering
  \small
\includegraphics[scale=1]{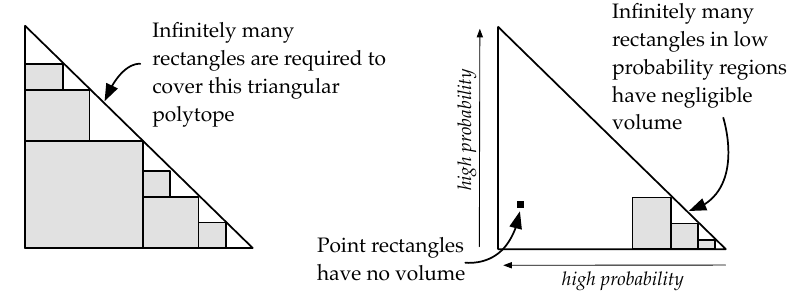}

\hspace{0in}(a)\hspace{1.7in}(b)
\caption{$\mathds{R}^2$ view of (a) hyperrectangular decomposition
and (b) hyperrectangle sampling, where density is concentrated
in the top-left corner}
\label{fig:decomp}
\end{figure}

Ideally, we would take a formula $\varphi$
and rewrite it as a disjunction of hyperrectangles $\bigvee H$,
but this disjunction is most likely an infinite one.
To see why this is the case, consider the simple
formula representing a triangular polytope
in Figure~\ref{fig:decomp}(a).
Here, there is no finite number of rectangles whose union is
the full region in $\mathds{R}^2$ enclosed by the triangle.

While the number of hyperrectangles enclosed in $\varphi$
is infinite, we can characterize them symbolically
using universal quantifiers, as shown by Li et al.~\cite{li2014management}.
Specifically, we define the hyperrectangular
decomposition of $\varphi$ as follows:

\begin{definition}[Hyperrectangular decomposition]
  \label{def:decomp}
  Given a formula $\varphi$, its \emph{hyperrectangular
  decomposition} $\rect_\varphi$ is:
  $$\rect_\varphi \equiv \biggl(\bigwedge_{x \in \vars_\varphi} l_x \leq u_x\biggr) \land
    \forall \vars_\varphi \ldotp
      \biggl(\bigwedge_{x \in \vars_\varphi}
        l_x \leq x \leq u_x\biggr) \Rightarrow \varphi$$
  where $l_x,u_x$ are fresh free variables introduced for each $x \in \vars_\varphi$,
  and $\forall \vars_\varphi$ is short for $\forall x_1,\ldots,x_n$,
  for $x_i \in \vars_\varphi$.

Given a model $m \models \rect_\varphi$, we say
that $H^m$ is the \emph{hyperrectangle induced by} $m$,
as defined below:
$$H^m \equiv \bigwedge_{x\in \vars_\varphi} m(l_x) \leq x \leq m(u_x)$$
\end{definition}

Intuitively, $\rect_\varphi$ characterizes every possible
hyperrectangle that is subsumed by $\varphi$.
The idea is that the hyperrectangle $H^m$ induced by each model
$m$ of $\rect_\varphi$ is 
%
subsumed by $\varphi$,
that is, $H^m \Rightarrow \varphi$.
The following example illustrates this process.

\begin{figure}[t]
    \centering
\includegraphics[scale=1]{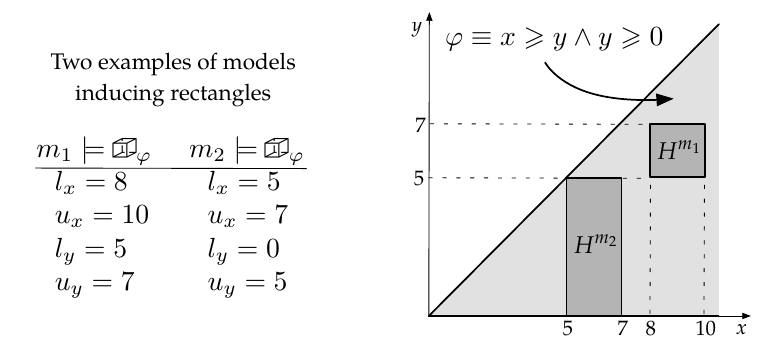}

\caption{Illustration of models of $\rect_\varphi$}
\label{fig:rectangle}
\end{figure}

\begin{example}
  Consider the formula
  $\varphi \equiv x \geq y \land y \geq 0$,
  illustrated in Figure~\ref{fig:rectangle}
  as a gray, unbounded polyhedron.
  The formula $\rect_\varphi$, after eliminating the universal quantifier,
  is:
  $$l_x \leq u_x \land l_y \leq u_y \land l_y \geq 0 \land l_x \geq u_y$$
  Figure~\ref{fig:rectangle} shows two models $m_1,m_2 \models \rect_\varphi$
  and their graphical representation  as rectangles $H^{m_1},H^{m_2}$ in $\mathds{R}^2$. Observe that both rectangles are subsumed by $\varphi$.
\end{example}

%

\paragraph{Hyperrectangle sampling}
Our symbolic weighted volume computation algorithm, $\volalg$,
is shown in Figure~\ref{alg:vol} as two
transition rules.
Given a pair $(\varphi, \pdfs)$,
the algorithm maintains a state
consisting of two variables: \rone $\volume$,
the current lower bound of the weighted
volume, and \rtwo $\Psi$, a constraint that encodes the \emph{remaining}
rectangles in the hyperrectangular decomposition of $\varphi$.

The algorithm is presented as guarded rules.
Initially, using the rule $\hdecomp$, $\volume$
is set to $0$ and $\Psi$ is set to $\rect_\varphi$.
The algorithm then proceeds by iteratively
applying the rule \sample.
Informally, the rule $\sample$ is used to
find arbitrary hyperrectangles in $\varphi$
and compute their weighted volume.
Specifically,
\sample finds a model $m$ of $\Psi$,
computes the weighted  volume of the hyperrectangle $H^m$ induced
by $m$, and adds the result to $\volume$.

To maintain soundness,
$\sample$ ensures that it never samples
two overlapping hyperrectangles, as otherwise
we would overapproximate the volume.
To do so, every time a hyperrectangle $H^m$ is sampled,
we conjoin an additional constraint to $\Psi$---denoted $\block(H^m)$
and defined in Figure~\ref{alg:vol}---%
that ensures that for all models $m' \models\Psi$,
$H^{m'}$ does not  overlap with $H^m$, i.e., $H^{m'} \land H^{m}$
is unsatisfiable.
Informally, the $\block(H^m)$ constraint specifies
that any newly sampled hyperrectangle should be to
the \emph{left} or \emph{right} of $H^m$ for at least one of the dimensions.

%

\paragraph{Lower and upper bounds}
The following theorem states the soundness of $\volalg$:
it maintains a lower bound on the exact
weighted volume.

\begin{theorem}[Soundness of $\volalg$]\label{thm:volsound}
The following is an invariant of
$\volalg(\varphi,\pdfs)$: $\volume \leq \vol(\varphi,\pdfs)$.
\end{theorem}

It follows from the above theorem that
we can use $\volalg$ to compute an upper bound
on the exact volume.
Specifically, because we are integrating over
\abrs{PDF}, we know that
$\vol(\varphi, \pdfs) + \vol(\neg \varphi,\pdfs) = 1$.
Therefore, by using \volalg to compute the weighted volume
of $\neg\varphi$, we get an upper bound
on the exact volume of $\varphi$.

\begin{corollary}[Computing upper bounds]
  \label{col:upper}
The following is an invariant of
$\volalg(\neg \varphi,\pdfs)$: $1 - \volume \geq \vol(\varphi,\pdfs)$
\end{corollary}

\subsection{Density-directed Sampling}
\label{sec:density}
While the \volalg algorithm is sound, it provides
no progress guarantees. Consider, for example,
a run that only samples unit hyperrectangles, i.e., points in $\mathds{R}^n$;
the volume of a point is 0, therefore, we will never compute
any volume.
Alternatively, the algorithm might diverge
by sampling hyperrectangles
in $\varphi$ that appear in very low probability density regions.
These two scenarios are illustrated in Figure~\ref{fig:decomp}(b)
on a triangular polytope in $\mathds{R}^2$.

Ideally, the rule \sample  would always find a model $m$ yielding
the hyperrectangle $H^m$ with the \emph{largest  weighted volume}.
Finding such a model amounts to solving the optimization
problem:
\begin{align*}
    \textstyle \underset{m \models \Psi}{\text{arg max}} \prod_{x_i \in \vars_\varphi}
    \int_{H^m_l(x_i)}^{H^m_u(x_i)}
    \pdf_i(x_i)~ dx_i
\end{align*}
From a practical perspective, there are no known tools
or techniques for finding models of first-order
formulas that maximize such complex objective functions, with integrals
over arbitrary probability density functions.

\begin{figure}[t]
  \centering
\begin{overpic}[scale=1.05]{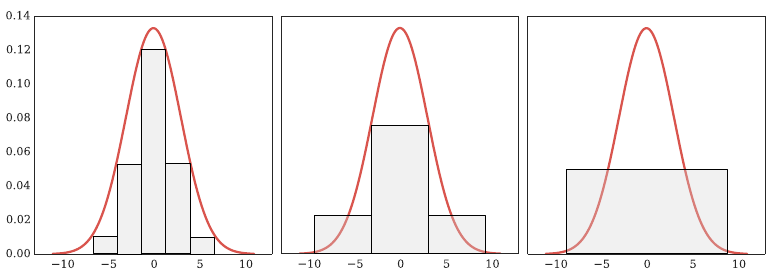}
    \put(30,30) {\footnotesize (a)}
    \put(62,30) {\footnotesize (b)}
    \put(94,30) {\footnotesize (c)}
\end{overpic}
%

\caption{Three \abrs{ADF} (gray) of a Gaussian \abr{PDF} (red) with mean 0 and
a standard deviation 3.
\abrs{ADF} are: (a) \emph{fine-grained};
(b) \emph{coarse}; (c) \emph{uniform}.}
\label{fig:approx}
\end{figure}

However, we make the key observation that
if $\pdf(x)$ is a \emph{step function}---i.e., piecewise constant---%
then we can symbolically encode the integral $\int \pdf(x)~dx$
in linear arithmetic.
As such, we propose to \rone \emph{approximate}
each density function $\pdf(x)$ with a step function $\appr(x)$,
\rtwo \emph{encode} the integrals $\int \appr(x)~dx$
as linear arithmetic formulas,
and \rthree \emph{direct} sampling towards hyperrectangles
that maximize these integrals, thus finding hyperrectangles
of large volume.

\paragraph{Approximate density functions and linear encodings}
We begin by defining \emph{approximate density functions} (\abrs{ADF}).

\begin{definition}[Approximate density functions]
    An approximate density function  $\appr(x)$
    is of the following form:
 \[
    \appr(x) = \begin{cases}
        c_i, &  x \in [a_i,b_i) ~~~\text{for}~~~ 1 \leq i \leq n\\
        0, & \text{otherwise}
      \end{cases}
  \]
where 
$c_i,a_i,b_i \in \mathds{R}$, $c_i > 0$,
and all $[a_i,b_i)$ are disjoint.
\end{definition}

We now show how to encode a formula $\sappr(x)$
over  the free variables $\delta_x, l_x, u_x$,
where for any model $m \models \sappr(x)$,
the value
 $m(\delta_x)$ is the area under $\appr(x)$
between $m(l_x)$ and $m(u_x)$, i.e.:
$m(\delta_x) = \int_{m(l_x)}^{m(u_x)} \appr(x)~dx$.
Intuitively, the value of this integral
is the sum of the areas of each \emph{bar} in $\appr(x)$,
restricted to
$[m(l_x), m(u_x)]$.


\begin{definition}[Encoding area under an \abr{ADF}]
Given an $\abr{ADF}$ $\appr(x)$, we define $\sappr(x)$ as follows:
\[
    \textstyle \sappr(x) \equiv \delta_x = \sum_{i = 1}^n c_i \cdot \bigl|[a_i,b_i) \cap [l_x,u_x]\bigr|
\]
\end{definition}

The  constraint $\sappr(x)$ is directly expressible
in linear arithmetic,
since
\[
\bigl|[a_i,b_i) \cap [l_x,u_x]\bigr| = \text{max}(\text{min}(b_i,u_x) - \text{max}(a_i,l_x),0)
\]
The finite sum in $\sappr(x)$ computes the size
of the intersection of $[l_x,u_x]$ with each
interval $[a_i,b_i)$ in  $\appr(x)$, and multiplies
the intersection with $c_i$, the value of the $\appr$ in that interval.
%

\begin{figure}[t!]
  \footnotesize
    \centering
\begin{prooftree}
\justifies
\volume \gets 0
\quad \Psi \gets \rect_\varphi
\quad \emph{lb} \gets 1
\using \hdecomp
\end{prooftree}

\vspace{.2in}
\begin{prooftree}
    m \models \Psi \land \bigwedge_{x_i\in \vars_\varphi}\exists\delta_{x_i} \ldotp \sappr_i(x_i) \land \delta_{x_i} \geq \emph{lb}
    \justifies
    \volume \gets \volume +\vol(H^m,\pdfs)
    \quad\quad \Psi \gets \Psi \land \block(H^m)
    \using \sample
\end{prooftree}

\vspace{.1in}
\begin{prooftree}
    \justifies
    \emph{lb} \gets \lambda*\emph{lb}
    \using \decay
\end{prooftree}

\caption{\volopt: \abr{ADF}-directed volume computation}
\label{fig:algadf}
\end{figure}

\paragraph{\abr{ADF}-directed Volume Computation}
We now present the algorithm \volopt (Figure~\ref{fig:algadf}), an extension of our volume computation
algorithm \volalg that uses \abrs{ADF} to steer the sampling process.
The \abrs{ADF} are only used for guiding the rule
\sample towards dense hyperrectangles, and thus do not affect
soundness of the volume computation.
For example, Figure~\ref{fig:approx} shows three approximations
of a Gaussian; all three are valid approximations.
In Section~\ref{sec:impl}, we discuss the impact
of different \abrs{ADF} on performance.

Formally, we create a set of \abrs{ADF} $\apprs = \{\appr_1,\ldots,\appr_n\}$,
where, for each variable $x_i \in \vars_\varphi$,
we  associate the \abr{ADF} $\appr_i(x_i)$.
The rule \sample now encodes $\sappr_i(x_i)$ and attempts
to find a hyperrectangle such that for each dimension $x$,
$\delta_x$ is greater than some lower bound $\emph{lb}$,
which is initialized to $1$.
Of course, we need to keep reducing the value $\emph{lb}$ as we run
out of hyperrectangles of a given volume.
Therefore, the rule \decay is used to shrink $\emph{lb}$ using a
fixed \emph{decay rate} $\lambda \in (0,1)$
and can be applied when \sample fails to find
a sufficiently large hyperrectangle.

Note that, ideally, we would look for a model $m$
such that $\prod_{x\in \vars_\varphi} \delta_x$ is maximized,
thus, finding the hyperrectangle with the largest weighted volume
with respect to the \abrs{ADF}.
However, this constraint is non-linear.
To lower the complexity of the problem
to that of linear arithmetic, we set a decaying lower bound
and attempt to find a model where each $\delta_x$
is greater than the lower bound.

\subsection{Convergence of \volopt}
We now discuss the convergence properties of $\volopt$.
Suppose we are given a  formula  $\varphi$, a set $\pdfs$,
and a set $\apprs$.
Let $R \subset \mathds{R}^n$ be the region where all the \abrs{ADF} in $\apprs$
are non-zero.
We will show that $\volopt$ converges, in the limit, to the
exact weighted volume restricted to $R$;
that is, $\volopt$ converges to
$\textstyle \int_{\varphi \cap R} \prod_{x_i\in \vars_\varphi} \pdf_i(x_i)~ d\vars_\varphi$.

Note the fascinating part here is that we do not impose
any restrictions on the \abrs{ADF}:
they do not have to
have any correspondence with the \abrs{PDF}
they approximate;
they need only
be step functions.
Of course, in practice,
the quality of the approximation dictates
the rate of convergence,
but we delay this discussion to Section~\ref{sec:impl}.

The following theorem states convergence of
$\volopt$; it assumes that \sample is applied
iteratively and \decay is only
applied when \sample cannot find a model.

\begin{theorem}[Convergence to $R$]\label{thm:limit}
    Assume $\volopt$ is run on $(\varphi,\pdfs)$
    and a set of \abrs{ADF}  $\apprs$
that are non-zero for $R \subset \mathds{R}^n$.
Let $\volume_i$ be the value of $\volume$ after
$i$ applications of \sample. Then,
$\textstyle \lim_{i \rightarrow \infty} \volume_i =  \int_{\varphi \cap R} \prod_{x_i\in \vars_\varphi} \pdf_i(x_i)~ d\vars_\varphi$.
\end{theorem}

%
Note that the above theorem
directly gives us a way to approach the exact volume.
Specifically, by performing runs of $\volopt$
on subsets in an infinite partition of $\mathds{R}^n$
induced by the \abrs{ADF},
we can ensure that the sum over the
$\volopt$ processes approaches the exact volume in the limit.
For all $i$, let $\apprs_i$ be a set of \abrs{ADF}
corresponding to an $\volopt$ process $P_i$,
where $R_i \subset \mathds{R}^n$ is the non-zero
region of $\apprs_i$.
We require an infinite set of $P_i$ to partition $\mathds{R}^n$:
\rone for all $i \neq j$, $R_i \cap R_j = \emptyset$, and
\rtwo $\bigcup_{i=1}^\infty R_i = \mathds{R}^n$.
The following theorem formalizes the argument:

\begin{theorem}[Convergence]\label{thm:limitfull}
  Let $P_1,P_2,\ldots$ be $\volopt$ processes
  that partition $\mathds{R}^n$.
  Assume a fair serialization
  where each $P_i$ performs $\sample$
  infinitely often, and
  let $\volume_n$ be the total computed volume
  across all $P_i$ after $n$ successful calls to $\sample$.
  Then, $\lim_{n \rightarrow \infty} \volume_n = \vol(\varphi,\pdfs)$.
\end{theorem}


\section{Implementation and Evaluation}\label{sec:impl}


\newcommand{\dt}[1]{\abr{DT}$_{#1}$}
\newcommand{\dta}[1]{\abr{DT}$_{#1}^{\alpha}$}
\newcommand{\svm}[1]{\abr{SVM}$_{#1}$}
\newcommand{\svma}[1]{\abr{SVM}$_{#1}^{\alpha}$}
\newcommand{\nn}[2]{\abr{NN}$_{#1,#2}$}


In this section, we present a case study on the
bias of decision-making programs.
In particular, we verify
the group fairness property with qualification, which requires
that a program $\prog$ satisfies the condition
\[
    \frac{\pr{\prog(\vec{v}) = \true  \mid \emph{min}(\vec{v}) = \true \land \emph{qual}(\vec{v}) = \true}}
    {\pr{\prog(\vec{v}) = \true  \mid \emph{min}(\vec{v}) = \false \land \emph{qual}(\vec{v}) = \true}}
>
0.85
\]
We implement our algorithms in a tool called \fs
and evaluate its ability to prove or disprove
the property for an assortment of programs;
furthermore, we evaluate
the effect of various parameters on the performance of the implementation,
and we compare the applicability of \fs to other probabilistic inference tools.

\subsection{Implementation}\label{ssec:impl}
We implemented our presented algorithms in a new tool
called \fs, which
employs  Z3~\cite{de2008z3} for \abr{SMT} solving and
Redlog~\cite{redlog} for quantifier elimination.
To compute the group fairness ratio, we decompose the conditional probabilities
into four joint probabilities:
$\pr{\prog(\vec{v}) \land \emph{min}(\vec{v}) \land \emph{qual}(\vec{v})}$,
$\pr{\prog(\vec{v}) \land \neg \emph{min}(\vec{v}) \land \emph{qual}(\vec{v})}$,
$\pr{\emph{min}(\vec{v}) \land \emph{qual}(\vec{v})}$, and
$\pr{\neg \emph{min}(\vec{v}) \land \emph{qual}(\vec{v})}$.
\fs computes lower bounds for eight quantities:
the weighted volume of each of these probabilities
and also their negations, since a lower bound
on the negation provides an upper bound on the positive form.
%
A \emph{round} of sampling involves \rone obtaining a
sample (hyperrectangle) for each of these eight quantities,
 \rtwo computing these samples' weighted volumes,
  \rthree
 updating the bounds on the group fairness ratio,
 and \rfour checking if bounds are precise enough
 to conclude \emph{fairness} or \emph{unfairness}.
Rounds of sampling are performed until a proof is found
or a timeout is reached.


\paragraph{Sample maximization}
A key optimization is the maximization of hyperrectangles
obtained during sampling.
We use Z3's optimization capability
 to maximize and minimize the finite bounds
of all hyperrectangles, while still
satisfying the formula $\Psi$ (in Figures~\ref{alg:vol} and~\ref{fig:algadf}).
This process is performed greedily by extending a hyperrectangle in
one dimension at a time to find a
maximal hyperrectangle.
If a dimension extends to infinity, then we drop that bound,
thus resulting in an unbounded hyperrectangle.

\subsection{Benchmarks}
We trained a variety of machine-learning models on a popular income dataset~\cite{dataset}
used in related research on algorithmic fairness~\cite{feldman15,zemel13,calders10}
to predict whether a person has a low or high income;
suppose, for example,
these programs would be used to determine
the salary of a new employee:
\emph{high} ($>\$50{,}000$) or \emph{low}.
We would like to verify whether
salary decisions are fair to qualified female employees.
Using the Weka machine learning suite~\cite{weka},
we learned 11 different decision-making programs
(see, e.g., Bishop's textbook~\cite{bishop2006pattern}
for background)
which are listed in Figure~\ref{fig:summary}:

\begin{itemize}
\setlength{\itemsep}{-2pt}
\item Four \emph{decision trees}, named \dt{n},
    where $n$ is the number of conditionals in the program.
    The number of variables and the depth of the tree
    each varies from 2 to 3.
\item Four \emph{support vector machines} with linear kernels, named \svm{n},
    where $n$ is the number of variables in the linear separator.
\item Three  \emph{neural networks} using \emph{rectified linear units}~\cite{nair2010rectified}, named \nn{n}{m},
    where $n$ is the number of input variables,
    and $m$ is the number of nodes in the single hidden layer.
\end{itemize}
%

As we will show in the next section,
some of these programs do not satisfy group fairness.
We introduced modifications of \dt{16} and \svm{4},
called \dta{16} and \svma{4},
that implement rudimentary forms of \emph{affirmative action}
for female applicants.
For \dta{16}, there is a 15\% chance it will flip a decision to give the low salary;
for \svma{4}, the linear separator is moved to increase the likelihood of hiring.

Additionally, we used three different probabilistic population models,
programs that define the probabilistic inputs, that were inferred from the same dataset:
\rone a set of \emph{independently distributed} variables,
\rtwo a \emph{Bayesian network} using a simple graph structure, and
\rthree the same Bayesian network,
but with an integrity constraint in the form of
an inequality between two of the variables.
%
Note that the first model is a trivial case:
since there is there is no dependence between variables,
all programs will be fair;
this simplicity serves well as a baseline for our evaluation.
The Bayesian models permit correlations between the variables,
allowing for more subtle sources of fairness or unfairness.
The benchmarks we use are derived from each combination
of population models with decision-making programs
\iftoggle{long}{
(see Appendix~\ref{app:code} for an example).
}
{
(see \emph{supplementary materials} for an example).
}

\subsection{Evaluation}

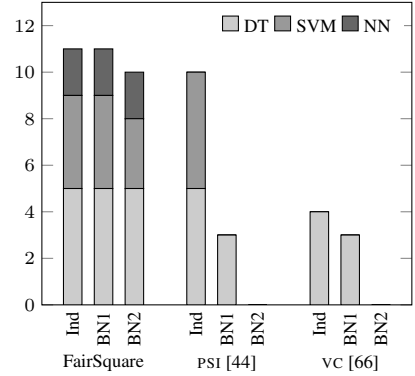
\begin{figure*}[t]
\centering
\begin{minipage}[b]{.6\textwidth}
\centering
\scriptsize
\setlength{\tabcolsep}{2.5pt}
\newcommand{\fair}[3]{\multicolumn{1}{c}{\cellcolor{green!20}\ding{51}} & { #1} & { #2} & { #3}}
\newcommand{\unfair}[3]{\multicolumn{1}{c}{\cellcolor{red!25}\ding{55}} & { #1} & { #2} & { #3}}
\newcommand{\stime}[4]{{\tiny $\overset{#1}{#2}$} & { #3} & { \abr{TO}} & { #4}}
\newcommand{\qtime}{\abr{TO}$_q$ & - & - & { \abr{TO}}}
\newcommand{\tblrow}[5]{#1 & #2 & #3 & #4 & #5}
\newcommand{\hhlinesep}{\hhline{|=##=##===##===##===|}}
\newcommand{\trih}{{ \it Res} & { \it \#} & { \it Vol} & { \it QE}}
\newcommand{\thline}{\specialrule{.15em}{.00em}{.00em}}
\renewcommand{\thline}{\hline}
{\def\arraystretch{1.0}
\begin{tabular}{cccccccccccccc}
    \toprule

    \multirow{3}{1.3cm}{\centering  \textbf{Decision program}} & \multirow{3}{*}{ \bf Acc} & \multicolumn{12}{c}{ \bf Population Model} \\
    & & \multicolumn{4}{c}{ Independent} & \multicolumn{4}{c}{ Bayes Net 1} &\multicolumn{4}{c}{ Bayes Net 2} \\
    \cmidrule(lr){3-6}
    \cmidrule(lr){7-10}
    \cmidrule(lr){11-14}
    & & \trih & \trih & \trih \\
    \midrule
    \tblrow{\dt{4}}{0.79}{\fair{10}{1.3}{0.5}}{\unfair{12}{2.2}{0.9}}{\unfair{18}{6.6}{2.2}} \\
    \tblrow{\dt{14}}{0.71}{\fair{20}{4.2}{1.4}}{\fair{38}{52.3}{11.4}}{\fair{73}{130.9}{33.6}} \\
    \tblrow{\dt{16}}{0.79}{\fair{21}{7.7}{2.0}}{\unfair{22}{15.3}{6.3}}{\unfair{22}{38.2}{14.3}} \\
    \tblrow{\dta{16}}{0.76}{\fair{18}{5.1}{3.0}}{\fair{34}{32.0}{8.2}}{\fair{40}{91.0}{19.4}} \\
    \tblrow{\dt{44}}{0.82}{\fair{55}{63.5}{9.8}}{\unfair{113}{178.9}{94.3}}{\unfair{406}{484.0}{222.4}} \\ \thline
    \tblrow{\svm{3}}{0.79}{\fair{10}{2.6}{0.6}}{\unfair{10}{3.7}{1.7}}{\unfair{10}{10.8}{6.2}} \\
    \tblrow{\svm{4}}{0.79}{\fair{10}{2.7}{0.8}}{\unfair{18}{13.3}{3.1}}{\unfair{14}{33.7}{20.1}} \\
    \tblrow{\svma{4}}{0.78}{\fair{10}{3.0}{0.8}}{\fair{22}{15.7}{3.2}}{\fair{14}{33.4}{63.2}} \\
    \tblrow{\svm{5}}{0.79}{\fair{10}{8.5}{1.3}}{\unfair{10}{12.2}{6.3}}{\qtime} \\
    \tblrow{\svm{6}}{0.79}{\stime{0.02}{35.3}{634}{2.4}}{\stime{0.09}{3.03}{434}{12.8}}{\qtime} \\ \thline
    \tblrow{\nn{2}{1}}{0.65}{\fair{78}{21.6}{0.8}}{\fair{466}{456.1}{3.4}}{\fair{154}{132.9}{7.2}}\\
    \tblrow{\nn{2}{2}}{0.67}{\fair{62}{27.8}{2.0}}{\fair{238}{236.5}{7.2}}{\fair{174}{233.5}{18.2}}\\
    \tblrow{\nn{3}{2}}{0.74}{\stime{0.03}{674.7}{442}{10.0}}{\stime{0.00}{5.24}{34}{55.9}}{\qtime} \\ \thline
\end{tabular}}
\end{minipage}%
\begin{minipage}[b]{.39\textwidth}
\centering
\pgfplotstableread{
    X   M    Tool DT  SVM NN
    1   Ind  F2   5   4   2
    2   BN1  F2   5   4   2
    3   BN2  F2   5   3   2
    5   Ind  \abr{PSI}  5   5   0
    6   BN1  \abr{PSI}  3   0   0
    7   BN2  \abr{PSI}  0   0   0
    9   Ind  \abr{VC}   4   0   0
    10  BN1  \abr{VC}   3   0   0
    11  BN2  \abr{VC}   0   0   0
}\datatable
\begin{tikzpicture}[baseline=(current bounding box.center)]
  \footnotesize
    \smaller
    \begin{axis}[
            width=6.5cm,
            xtick=data,
            xticklabels from table={\datatable}{M},
            xticklabel style={rotate=90,anchor=east},
            xtick style={draw=none},
            enlarge x limits=0.1,
            ymin=0,ymax=13,
            ytick={0,2,4,6,8,10,12},
            ybar stacked,
            bar width=7pt,
            legend style={draw=none,legend columns=-1},
            after end axis/.append code={
                \path
                (axis cs:2,0) node [yshift=-7ex] {\fs}
                (axis cs:6,0) node [yshift=-7ex] {\abr{PSI}~\cite{Gehr16}}
                (axis cs:10,0) node [yshift=-7ex] {\abr{VC}~\cite{sankaranarayanan13}};
            }
        ]
        \addplot[draw=black,fill=black!20] table[x=X,y=DT]\datatable; \addlegendentry{DT}
        \addplot[draw=black,fill=black!40] table[x=X,y=SVM]\datatable; \addlegendentry{SVM}
        \addplot[draw=black,fill=black!60] table[x=X,y=NN]\datatable; \addlegendentry{NN}
    \end{axis}
\end{tikzpicture}
\end{minipage}
\caption{
{\bf(Left)}
Results of \fs applied to 39 group fairness problems.
\emph{Res} indicates
\ding{51} for fair; \ding{55} for unfair.
\emph{Vol} indicates the time (s)
of the sampling procedure;
\# is the number of \abr{SMT} calls.
If sampling timed out (900s),
\emph{Res} denotes the latest bounds
on the fairness ratio.
\emph{QE} indicates the time (s)
of the quantifier elimination procedure used prior to sampling;
if this timed out (900s),
we could not perform any sampling,
denoted by a \abr{TO}$_q$ for \emph{Res}.
\emph{Acc} is the training set accuracy
for each of the programs.
{\bf(Right)}
Comparison of the number of benchmarks
that \fs, \abr{PSI}~\cite{Gehr16}, and \abr{\abr{VC}}~\cite{sankaranarayanan13} were able to
solve.
}
\label{fig:summary}
\end{figure*}


In this section, we discuss the ability of \fs
to verify qualified group fairness or unfairness for 39 problems,
as summarized in Figure~\ref{fig:summary}.
In these problems, $\emph{min}(\vec{v}) = \true$ when the applicant is female,
and $\emph{qual}(\vec{v}) = \true$ in two different scenarios:
first, we consider the case when $\emph{qual}$ is tautologically true,
and second, when the applicant is at least 18 years of age.
Figure~\ref{fig:summary} shows only the former case,
as while the numbers are different for the two cases,
the qualitative results are quite similar.
\iftoggle{long}{
See Appendix~\ref{app:res} for the full table of results.
}
{
See \emph{supplementary materials} for the full table of results.
}
We fix $\epsilon = 0.15$.
To guide volume computation,
all Gaussian distributions with mean $\mu$ and variance $\sigma^2$
use \abrs{ADF} with 5 equal-width steps spanning $(\mu - 3\sigma^2,\mu + 3\sigma^2)$---analogous to Figure~\ref{fig:approx}(a).
\fs was able to solve 32 of the 39 problems, proving 21 fair and 11 unfair,
as shown in the left table of Figure~\ref{fig:summary}.


Consider the results for \dt{4}:
\fs proved it fair with respect to the independent population model
after 0.5 seconds of an initial quantifier elimination procedure
and 1.3 seconds of the actual volume computation algorithm,
which required 14 \abr{SMT} queries.
The more sophisticated Bayesian network models took longer for sampling,
but due to the correlations between variables,
were proved unfair.

In contrast, consider the results for \dt{44}
under the Bayes Net 1 population model:
\fs was unable to conclude fairness or unfairness
after 900 seconds of volume computation
(denoted by \abr{TO} in the \emph{Vol} column).
The lower and upper bounds of the fairness ratio
it had computed at that time are listed in the \emph{Res} column:
in this case, the value of the fairness ratio
is within $[0.70,0.88]$, which is not
precise enough for the $\epsilon = 0.15$ requirement
(but would be precise enough for $\epsilon$
outside of $[0.12,0.30]$).

In general, all conclusive results using the independent population model
were proved to be fair, as expected,
but many are unfair with respect to the clusters and Bayes net models
because of the correlations they capture.
This difference illustrates the sensitivity of fairness
to the population model;
in particular, none of the decision trees
syntactically access \emph{sex},
yet several are unfair.

Figure~\ref{fig:summary} also shows that the modifications
for affirmative action in \dta{16} and \svma{4}
are sufficient to make the programs fair
with respect to all of our population models
without making a substantial impact on the training set accuracy.

In summary, \fs is powerful enough to reason about
group fairness for many non-trivial programs.


\subsection{Comparison to other tools}
We ran our benchmarks on the two other \emph{exact} probabilistic inference tools
that accept the same class of problems,
and report these results in the right bar graph in Figure~\ref{fig:summary}.
First, we compare to Sankaranarayanan et al.'s tool~\cite{sankaranarayanan13} (\abr{VC}),
which is algorithmically similar to our tool:
it finds bounds for probabilities on individual paths
by approximating formulas with bounding and inscribed hyperrectangles.
Second, we compare to \abr{PSI}~\cite{Gehr16},
which symbolically computes representations of
the posterior distributions of variables.
When closed form solutions to the \abrs{CDF} exist,
\abr{PSI} is a great aid because these solutions can
be evaluated to obtain exact values of probabilities instantly;
however, the graph illustrates that our benchmarks are often too complex,
resulting in integrals that do not have closed forms,
or problems that \abr{PSI} cannot solve within the timeout period.

Tools were deemed to have failed on a benchmark when they timed out
after a 900s period.
Additionally, the bounds on probability quantities from \abr{VC}
were often not tight enough for a proof.
\iftoggle{long}{
See Appendix~\ref{app:res} for the full quantitative table of results.
}
{
See \emph{supplementary materials} for the full quantitative table of results.
}

The figure illustrates some qualitative properties of the applicability
of the tools.
In general, all can solve most of the decision trees because
the trees partition the decision space using a number of inequalities
between a single variable and a \emph{constant}.
However, the presence of inequalities
involving multiple variables
can result in two phenomena:
\rone the lack of closed form posterior \abrs{CDF},
as reflected in the output of \abr{PSI}, and
\rtwo angled boundaries in the decision space
that are hard to approximate with hyperrectangles.
These inequalities occur in the \abrs{SVM},
neural networks, and the Bayes Net 2 population model.
Consequently,
\abr{VC} fails to produce good bounds in these cases;
\fs performs better, but also fails on the largest \abr{SVM}.

\subsection{Effect of Parameters}

\begin{figure}[t]
  \smaller
  \centering
    \setlength{\tabcolsep}{0pt}
    \begin{tabular}{rcc}
        & \dt{16} & \svm{4} \\
        \rotatebox{90}{\smaller \hspace{1em} \parbox{3cm}{\centering fairness ratio without sample maximization}} &
        \begin{overpic}[scale=0.25]{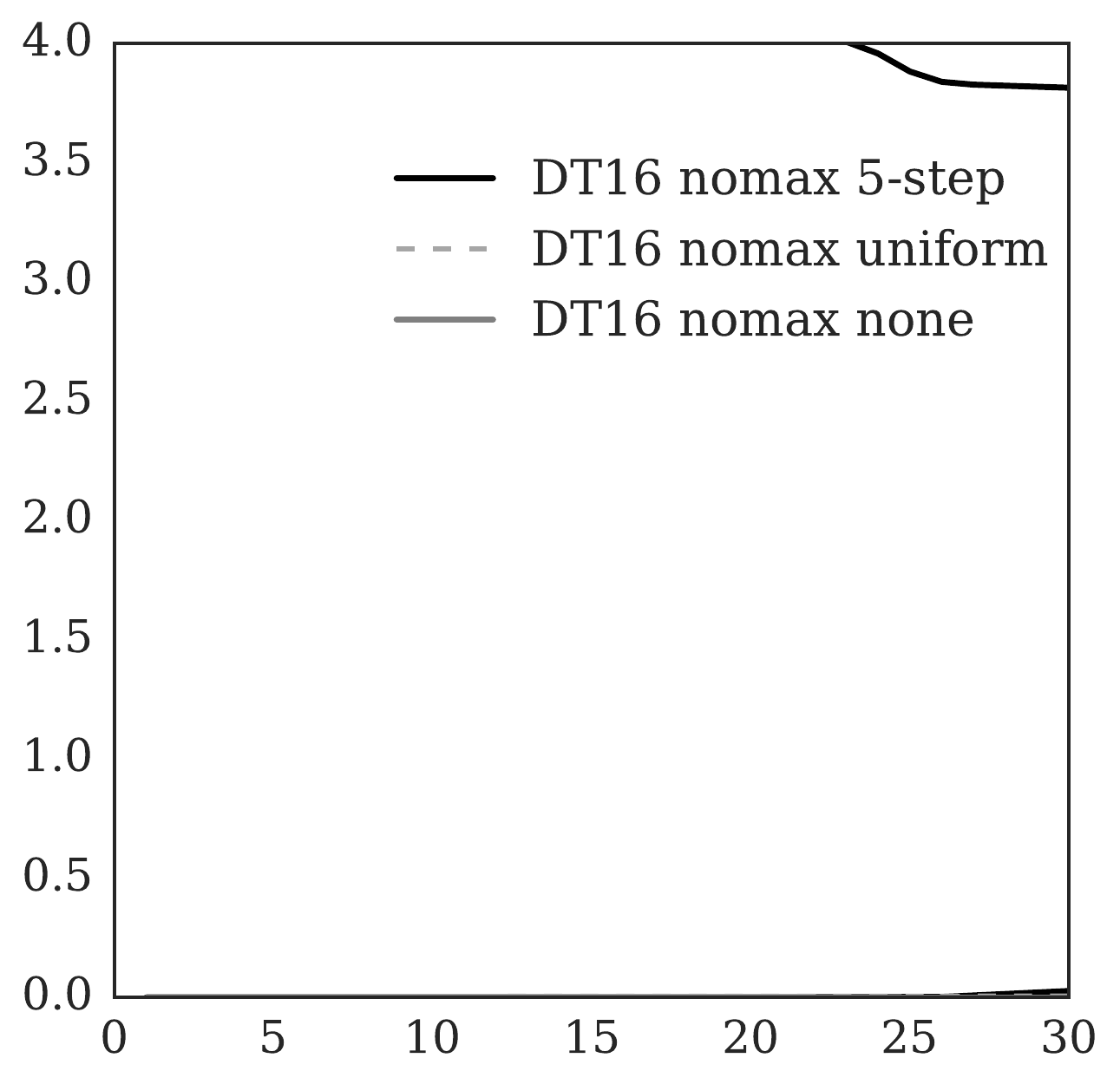} \put(80,50) {(a)} \end{overpic}&
        \begin{overpic}[scale=0.25]{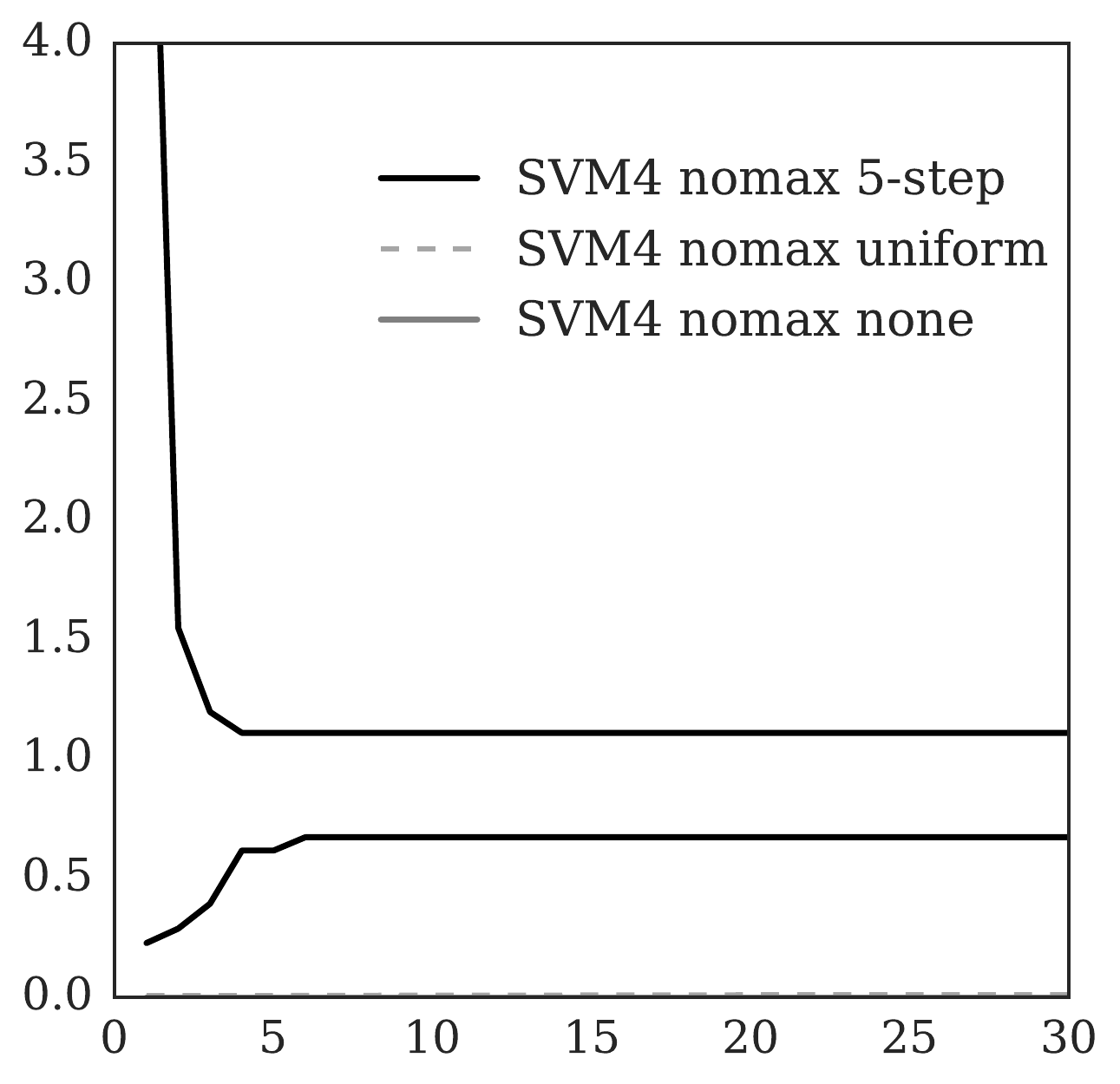} \put(80,50) {(c)} \end{overpic} \\
        \rotatebox{90}{\smaller \hspace{1.5em} \parbox{3cm}{\centering fairness ratio with sample maximization}} &
        \begin{overpic}[scale=0.25]{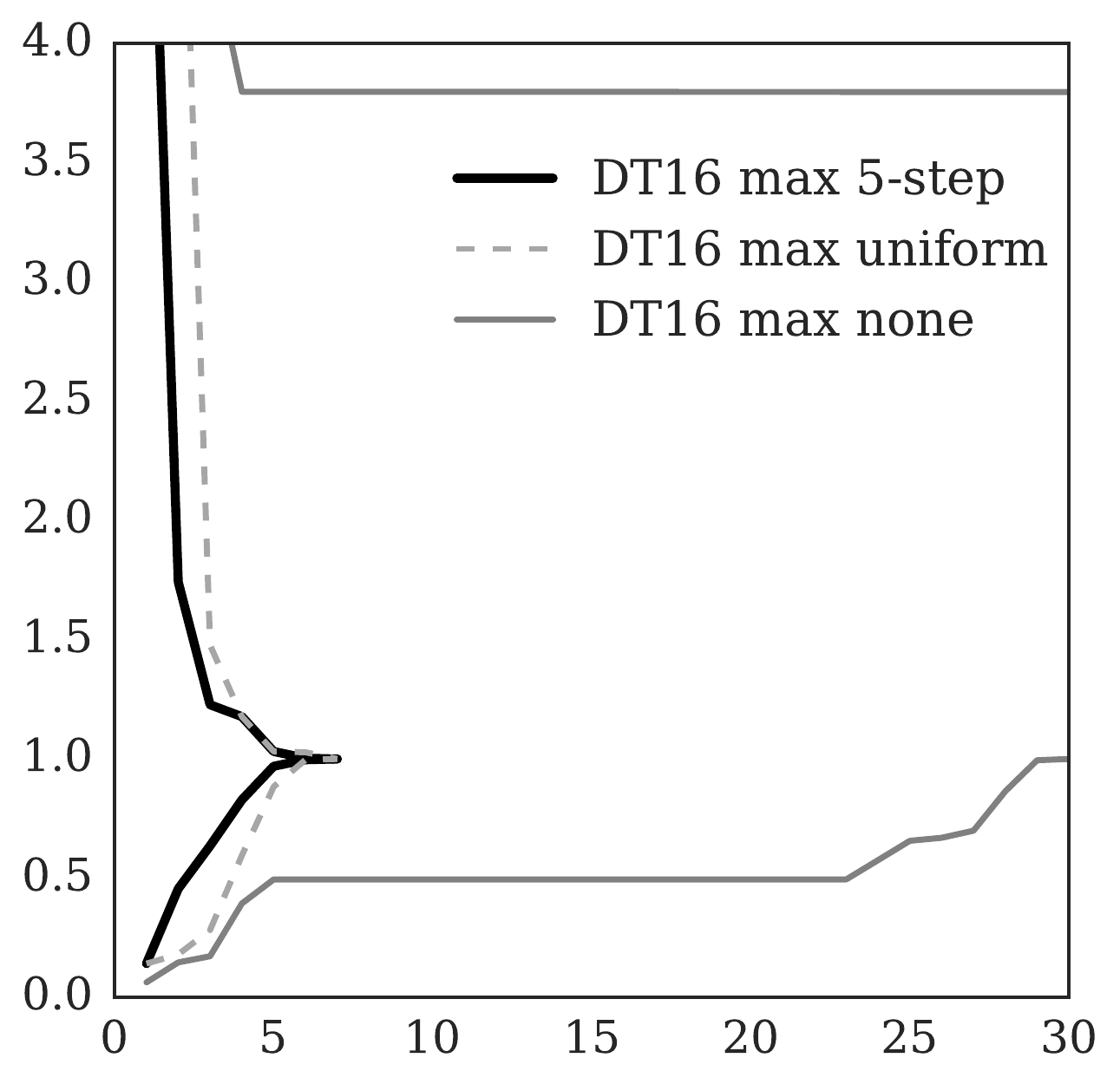} \put(80,50) {(b)} \end{overpic}&
        \begin{overpic}[scale=0.25]{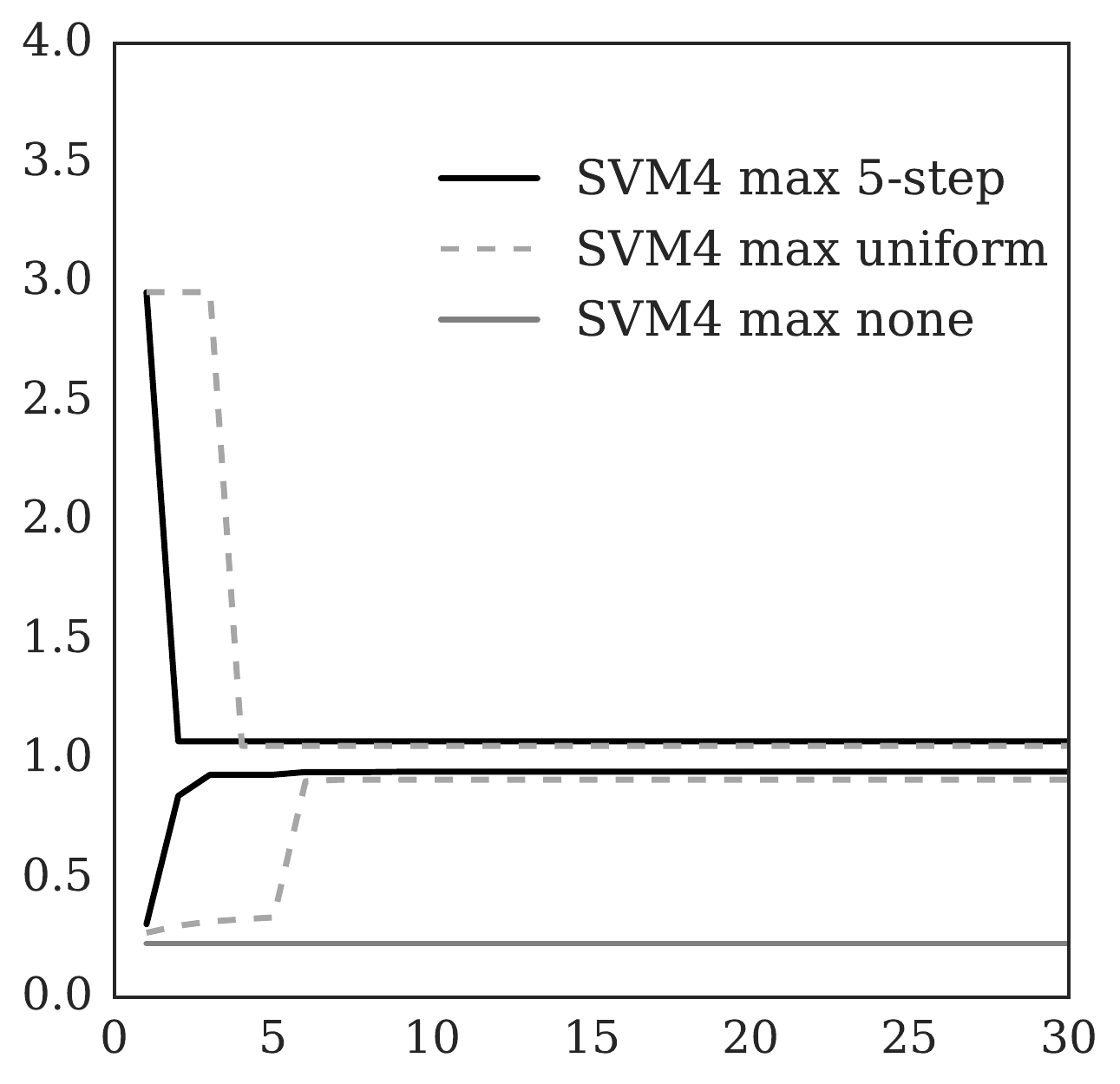} \put(80,50) {(d)} \end{overpic}
    \end{tabular}
\caption{
    Fairness ratio vs. rounds of sampling
    for \dt{16} and \svm{4}
    (independent pop model)
    differing on \abrs{adf}
    and sample maximization.
    In (b) two runs end at the exact value.
    Outside the visible range are:
    (a)(c) upper and lower bounds of \emph{uniform} and \emph{none};
    (d) upper bounds of \emph{none}.
}
\label{fig:opt}
\end{figure}

\begin{figure}[t]
    \centering
    \pgfplotsset{every tick label/.append style={font=\tiny}}
    \setlength{\tabcolsep}{0pt}
    \smaller
    \begin{tabular}{rcc}
        & ~~~~~~~~\dt{16} & ~~~~~~~~\svm{4} \vspace{0em} \\
        \rotatebox{90}{\smaller \hspace{1em} \parbox{2cm}{\centering average weighted volume per sample}} &
        \multicolumn{1}{r}{\begin{tikzpicture}
            \begin{axis}[
                    width=5cm,
                    height=3.2cm,
                    legend style={font=\tiny, legend columns=-1, at={(1,1.5)}},
                    ybar, ymin=0, ymax=0.15,
                    yticklabel style={
                        /pgf/number format/fixed,
                        /pgf/number format/precision=2
                    },
                    scaled y ticks=false,
                    symbolic x coords={5step,uniform,none},
                    xtick=data,
                    xticklabels={
                        5-step,
                        uniform,
                        none,
                    },
                    xticklabel style={text width=1cm, align=center},
                    bar width=7, x=1cm, enlarge x limits={abs=0.5cm},
                    after end axis/.append code={
                        \path (rel axis cs:0.9,0.85) node {(a)};
                    }
                    ]
                \addplot[ybar,fill=black] coordinates {
                    (5step, 0.1429)
                    (uniform, 0.1429)
                    (none, 0.0307)
                };
                \addplot[ybar,fill=black!20] coordinates {
                    (5step, 0.0236)
                    (uniform, 0.0123)
                    (none, 0.0022)
                };
                \legend{\dt{16} max, \dt{16} nomax}
            \end{axis}
        \end{tikzpicture}} &
            \multicolumn{1}{r}{\begin{tikzpicture}
            \begin{axis}[
                    width=5cm,
                    height=3.2cm,
                    legend style={font=\tiny, legend columns=-1, at={(1,1.5)}},
                    ybar, ymin=0, ymax=0.15,
                    symbolic x coords={5step,uniform,none},
                    yticklabel style={
                        /pgf/number format/fixed,
                        /pgf/number format/precision=2
                    },
                    scaled y ticks=false,
                    xtick=data,
                    xticklabels={
                        5-step,
                        uniform,
                        none,
                    },
                    xticklabel style={text width=1cm, align=center},
                    bar width=7, x=1cm, enlarge x limits={abs=0.5cm},
                    after end axis/.append code={
                        \path (rel axis cs:0.9,0.85) node {(c)};
                    }
                    ]
                \addplot[ybar,fill=black] coordinates {
                    (5step, 0.0331)
                    (uniform, 0.0330)
                    (none, 0.0206)
                };
                \addplot[ybar,fill=black!20] coordinates {
                    (5step, 0.0326)
                    (uniform, 0.0218)
                    (none, 0.0111)
                };
                \legend{\svm{4} max, \svm{4} nomax}
            \end{axis}
        \end{tikzpicture}}  \\
        \rotatebox{90}{\smaller \hspace{1.5em} \parbox{2cm}{\centering average time per round}} &
        \multicolumn{1}{r}{\begin{tikzpicture}
            \begin{axis}[
                    width=5cm,
                    height=3.2cm,
                    ybar, ymin=0, ymax=2,
                    symbolic x coords={5step,uniform,none},
                    xtick=data,
                    xticklabels={
                        5-step,
                        uniform,
                        none,
                    },
                    xticklabel style={text width=1cm, align=center},
                    bar width=7, x=1cm, enlarge x limits={abs=0.5cm},
                    after end axis/.append code={
                        \path (rel axis cs:0.9,0.85) node {(b)};
                    }
                    ]
                \addplot[ybar,fill=black] coordinates {
                    (5step, 0.9721)
                    (uniform, 0.8798)
                    (none, 0.6270)
                };
                \addplot[ybar,fill=black!20] coordinates {
                    (5step, 0.2491)
                    (uniform, 0.1924)
                    (none, 0.1908)
                };
            \end{axis}
        \end{tikzpicture}} &
        \multicolumn{1}{r}{\begin{tikzpicture}
            \begin{axis}[
                    width=5cm,
                    height=3.2cm,
                    ybar, ymin=0, ymax=2,
                    symbolic x coords={5step,uniform,none},
                    xtick=data,
                    xticklabels={
                        5-step,
                        uniform,
                        none,
                    },
                    xticklabel style={text width=1cm, align=center},
                    bar width=7, x=1cm, enlarge x limits={abs=0.5cm},
                    after end axis/.append code={
                        \path (rel axis cs:0.9,0.85) node {(d)};
                    }
                    ]
                \addplot[ybar,fill=black] coordinates {
                    (5step, 1.4835)
                    (uniform, 1.6445)
                    (none, 1.3091)
                };
                \addplot[ybar,fill=black!20] coordinates {
                    (5step, 0.3002)
                    (uniform, 0.2006)
                    (none, 0.2289)
                };
            \end{axis}
        \end{tikzpicture}} \\
    \end{tabular}
\caption{
    Effect of optimizations
    on  performance of \fs for
    \dt{16} and \svm{4} (independent pop model).
    (a) and (c) show the average weighted volume
    per sample (averaged across all computed probabilities).
    (b) and (d) show the average time (s)
    per round of sampling.
}
\label{fig:optbar}
\end{figure}
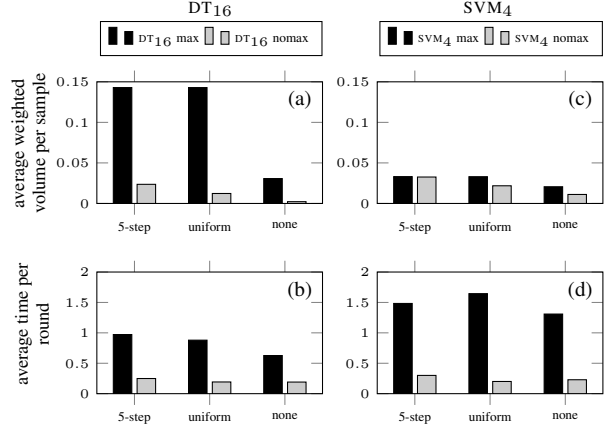

We now explore  the effects
of the approximate density functions (see Section~\ref{sec:density})
and of the sample maximization optimizations.
These results are captured in Figures~\ref{fig:opt} and~\ref{fig:optbar}.

There are three instances of \abrs{ADF} in Figure~\ref{fig:opt}
used to guide the sampling to high-probability regions:
\rone \emph{none} indicates that no \abr{ADF} is used, i.e.,
we used \volalg instead of \volopt;
\rtwo \emph{uniform} indicates that each $\texttt{gauss}(\mu, \sigma^2)$ is approximated by a
uniform function spanning $(\mu - 3\sigma^2, \mu + 3\sigma^2)$
(similar to Figure~\ref{fig:approx}(c)); and
\rthree \emph{5-step} indicates that each Gaussian is approximated by
a step function of 5 equal-width regions spanning that same domain
(similar to Figure~\ref{fig:approx}(a)).
Another variable, \emph{max} or \emph{nomax},
denotes whether the sample maximization optimization is enabled
(as described in Section~\ref{ssec:impl}).

Each combination of these techniques is run on two of our benchmarks:
\dt{16} and \svm{4} under the independent population model.
Figure~\ref{fig:opt}(a) and (c) show how
convergence to the fairness ratio is improved by
the choice of distribution approximation when sample maximization is not employed:
in particular, the runs using \emph{uniform} and \emph{none}
are not even visible, as the bounds never fall within $[0.01,4.0]$.
Plots (b) and (d) show that when sample maximization \emph{is} employed,
the choice between the uniform and 5-step approximations is
not as substantial on these benchmarks, although
\rone the better approximation gets better bounds faster, and
\rtwo using none results in substantially worse bounds.

Figure~\ref{fig:optbar} plot (a) and (c) show that employing \abrs{ADF}
and using sample maximization each increases the average weighted volume per sample,
which allows volume computation to be done with fewer samples.
Plots (b) and (d) illustrate the trade-off:
the average time per round of sampling
tends to be greater for more sophisticated optimizations.

We present these results for two particular problems
and observe the same results across our suite.
In summary, we have found that \abrs{ADF} and
sample maximization are both
 necessary for adequate performance of \fs.

\section{Related Work}\label{sec:relwork}
%

\paragraph{Probabilistic program analysis}
We refer the reader to Gordon et al. for a thorough survey~\cite{gordon2014probabilistic}.
%
A number of works tackled analysis of probabilistic programs
from an abstract interpretation perspective~\cite{monniaux2001backwards,monniaux2000abstract,monniaux2001abstract,mardziel2011dynamic,claret2013bayesian}.
The comparison between our solution through volume
computation and abstract interpretation is perhaps
analogous to \abr{SMT} solving and software model checking
versus abstract interpretation.
For example, techniques like Monniaux's~\cite{monniaux2000abstract}, sacrifice precision
(through joins, abstraction, etc.)
of the analysis for the benefit of efficiency.
Our approach, on the other hand, is aimed at eventually
producing a proof, or iteratively improving
probability bounds while guaranteeing convergence.


A number of works have also addressed probabilistic
analysis through symbolic execution~\cite{filieri2013reliability,sankaranarayanan13,geldenhuys2012probabilistic,sampson2014expressing}.
Filieri et al.~\cite{filieri2013reliability}
and Geldenhuys et al.~\cite{geldenhuys2012probabilistic}
attempt to find the probability
a safety invariant is preserved.
Both methods reduce to a weighted model counting approach
and are thus effectively restricted to variables over finite domains.
Note that our technique is more general than a  model counting approach,
as we can handle the discrete cases with a proper encoding
of the variables into a continuous domain.

A number of works rely on \emph{sampling} to approximate
the probability of a given program property. The Church~\cite{goodman08}
programming language, for instance, employs the
\emph{Metropolis--Hastings} algorithm~\cite{chib1995understanding},
a \emph{Markov Chain Monte Carlo} (\abr{MCMC}) technique.
Other  techniques perform probabilistic inference
by compiling programs or program paths
to Bayesian networks~\cite{koller2009probabilistic}
and applying \emph{belief propagation}~\cite{minka2012infer} or sampling~\cite{sampson2014expressing} on the network.



\paragraph{Volume computation}
The computation of weighted volume is known to be hard---even
for a polytope, volume computation is \#P-hard~\cite{khachiyan1993complexity}.
Two general approaches exist:
approximate and exact solutions.
Note that in general, any approximate technique
at best can prove facts \emph{with high probability}.

Our volume computation algorithm is inspired
by \rone Li et al.'s~\cite{li2014management}
formula decomposition procedure,
where quantifier elimination is  used
to under-approximate an \abr{LRA}
constraint as a Boolean combination of monadic
predicates;
and \rtwo Sankaranarayanan et al.'s~\cite{sankaranarayanan13}
technique for bounding the weighted
volume of a polyhedron,
which is the closest volume computation work to ours.
(The general technique of approximating complex regions
with unions of orthogonal polyhedra is well-studied
in hybrid systems literature~\cite{Bournez99}.)

A number of factors differentiate our work from~\cite{sankaranarayanan13},
which we compared with experimentally in Section~\ref{sec:impl}.
First, our approach is more general, in that it
can operate on Boolean formulas over linear and polynomial
inequalities, as opposed to just conjunctions of linear inequalities.
Second, our approach employs approximate distributions to guide
the sampling of hyperrectangles with large volume, which, as we have demonstrated experimentally, is a crucial feature of our approach.
Third, we provide theoretical convergence guarantees.

\emph{LattE} is a tool that performs exact integration
of polynomial functions over polytopes~\cite{de2012software}.
Belle et al.~\cite{belle15,belle16}
compute the volume of a linear real arithmetic (\abr{LRA}) formula
by, effectively, decomposing it into \abr{DNF}---a set of polyhedra---%
and using LattE to compute the volume of each polyhedron
with respect to piece-wise polynomial densities.
Our volume computation algorithm is more general in that
it \rone  handles formulas over real closed fields, which subsumes \abr{LRA},
and \rtwo handles arbitrary probability distributions.

Chistikov et al.~\cite{chistikov14arxiv,chistikov15}
present a framework for approximate counting with probabilistic
guarantees in \abr{SMT} theories,
which they specialize for bounded \abr{LRA}.
In contrast, our technique
\rone handles unbounded formulas in \abr{LRA}
as well as real closed fields,
\rtwo handles arbitrary distributions,
and \rthree provides converging lower-bound guarantees.
It is important to note
that there is a also a rich body of work
investigating randomized polynomial algorithms
for approximating the volume of a polytope,
beginning with Dyer et al.'s seminal work~\cite{dyer1991random}
(see Vempala~\cite{vempala2005geometric}
for a survey).

\paragraph{Algorithmic fairness}
Our work is inspired by recent concern in the
fairness of modern decision-making programs~\cite{zarsky2014understanding,barocas2014big}.
A number of recent works have explored algorithmic fairness~\cite{zemel13,feldman15,hardt16,dwork12,calders10,pedreshi2008discrimination,datta2016algorithmic,datta2015automated}.
For instance, Zemel et al.~\cite{zemel13} and Feldman et al.~\cite{feldman15}
study fairness from a machine learning classification perspective,
e.g., automatically learning fair classifiers.
Both works operate with a notion of fairness on a provided data set;
in contrast, we prove fairness with respect to a given probabilistic model of the population (a data set can be viewed as a special case).
%

Discrimination in \emph{black-box} systems
has been studied through the lens of statistical analysis~\cite{sweeney2013discrimination,datta2015automated,datta2016algorithmic}.
Notably, Datta et al.~\cite{datta2015automated}
created an automated tool that analyzes online advertising:
it operates dynamically by surveying the ads produced by Google.
Our approach differs from the statistical analyses
in that we require transparency of the decision procedure,
instead of data on the results of the decision procedure.

\paragraph{Future work}
We presented a new technique for verification
of probabilistic programs and applied to
quantifying bias of decision-making programs.
Our goals for future work include
exploring the applicability of our technique
to other notions of fairness properties (e.g.,~\cite{dwork12});
further refinements to the use of \abrs{ADF},
e.g. dynamically adapting them during \volopt;
and improving the weighted volume computation algorithm
for better scalability.




\bibliographystyle{abbrvnat}
\bibliography{biblio}

\appendix
\iftoggle{long}{
\section{Operational Semantics}\label{app:opsem}
\paragraph{Operational semantics}
We now define the operational semantics
of our program model.
Typically,
the \emph{state} $\val : V \rightarrow \mathds{R}$  of the program is defined
as a valuation function from variables in $V$ to
values in $\mathds{R}$.
In a probabilistic setting, however, we need to maintain
an additional state that dictates values drawn from probability distributions.
Following standard semantics of probabilistic programs~\cite{kozen1981semantics},
we assume a finite sequence of independent and identically
distributed (\emph{iid}) random variables.
The semantics of an execution is thus defined for a
fixed sequence of values $\omega$ of these variables.
Informally, it is as if we performed all sampling
before the program executes and stored the results
in a sequence for use whenever we encounter a probabilistic
assignment.
For simplicity of exposition, we assume all probabilistic
assignments sample from the same distribution; otherwise,
multiple sequences can be used, one per distribution.

We can now define the semantics of $\prog$ as shown in Figure~\ref{fig:sem}.
We use $\val(e)$ to denote the value of expression
$e$ under a given state $\val$, and use $\skips$
to denote the empty statement sequence.
We use the substitution notation
$s[x \mapsto c]$ to denote the state $s$ but with
$x$ mapped to $c \in \mathds{R}$.
The interesting rule is \passign, which, given a probabilistic
assignment $x \sim \pdf$, for some variable $x$ and distribution $\pdf$,
picks and removes the first element $c$ of the sequence $c:\omega$
to update the value of the variable $x$.

We assume no variable is used before being assigned to.
In a closed program,
we assume all variables are initially assigned to $0$.
Given a closed program $\prog$, we define $\sem{\prog}_\omega$
as the final state  $s$ reachable from executing $\prog$
using the sequence $\omega$, as defined by the
relation $\rightarrow$ in Figure~\ref{fig:sem}.

\begin{figure*}[t]
  \centering
  \smaller
  \begin{prooftree}
  \val' = \val[x \mapsto \val(e)]
  \justifies
  (\omega, \val, x \gets e) \rightarrow (\omega, \val', \skips)
  \using \assign
  \end{prooftree}
  \hspace{.5in}
  \begin{prooftree}
  \val' = \val[x \mapsto c]
  \justifies
  (c:\omega, \val, x \sim p) \rightarrow (\omega, \val', \skips)
  \using \passign
  \end{prooftree}
  \hspace{.5in}
  \begin{prooftree}
  (\omega, \val, S_1) \rightarrow (\omega', \val', S_1')
  \justifies
  (\omega, \val, S_1S_2) \rightarrow (\omega', \val', S_1'S_2)
  \using \seq
  \end{prooftree}
  ~\\
  \vspace{.1in}

  \begin{prooftree}
    ~\vspace{.085in}
  \justifies
  (\omega, \val, \skips S) \rightarrow (\omega, \val, S)
  \using \textsc{skip}
  \end{prooftree}
  \hspace{.4in}
  \begin{prooftree}
  \val(b) = \true
  \justifies
  (\omega, \val, \sif~b~\sthen~S_1~\selse~S_2) \rightarrow (\omega, \val, S_1)
  \using \tcond
  \end{prooftree}
  \hspace{.4in}
  \begin{prooftree}
  \val(b) = \false
  \justifies
  (\omega, \val, \sif~b~\sthen~S_1~\selse~S_2) \rightarrow (\omega, \val, S_2)
  \using \fcond
  \end{prooftree}
  \caption{Operational semantics}
  \label{fig:sem}
\end{figure*}

\section{Proofs} \label{app:proofs}

In this section, we prove correctness of the various
pieces of our algorithm. First, we introduce preliminary theorems.

The following theorem states the soundness and completeness of hyperrectangular
decomposition: models of $\rect_\varphi$ characterize all hyperrectangles in $\varphi$
and no others.

\begin{theorem}[Correctness of $\rect$]\label{thm:decomp}
Let $\varphi \in \smt$.
  \textbf{Soundness}:
Let $m \models \rect_\varphi$. Then, $H^m \Rightarrow \varphi$
is valid.
\textbf{Completeness}:
Let $H$ be a hyperrectangle such that  $H \Rightarrow \varphi$.
Then, the following is satisfiable:
$\rect_\varphi \land \bigwedge_{x \in \vars_\varphi} l_x = H_l(x) \land u_x = H_u(x) $
\end{theorem}

The following theorem  states the correctness of $\block$:
 it removes all hyperrectangles that overlap with $H^m$
 (soundness),
and it does not overconstrain $\Psi$ by removing hyperrectangles
that do not overlap with $H^m$ (completeness).

\begin{theorem}[Correctness of $\block$] \label{thm:block}
Given $\varphi$, let $\Psi \Rightarrow \rect_\varphi$, and
let $m_1,m_2 \models \Psi$.
%
\textbf{Soundness}:
    If $H^{m_1} \land H^{m_2}$ is satisfiable,
    then $m_2 \not\models \Psi \land \block(H^{m_1})$.
%
    \textbf{Completeness}: If $H^{m_1} \land H^{m_2}$
is unsatisfiable,
then $m_2 \models \Psi \land \block(H^{m_1})$.
\end{theorem}

The following theorem states the correctness of the \abr{ADF}
encoding:

\begin{theorem}[Correctness of $\sappr$]\label{thm:adf}
Fix an \abr{ADF} $\appr(x)$.
  \textbf{Soundness}:
For any model $m \models \sappr$,
the following  is true:
$m(\delta_x) = \int_{m(l_x)}^{m(u_x)} \appr(x)~dx$.
\textbf{Completeness}:
For any constants $a,b,c\in \mathds{R}$ such that
$c = \int_a^b \appr(x)~dx$, the following formula is satisfiable:
$\delta_x = c \land l_x = a \land u_x = b \land \sappr(x)$.
\end{theorem}

\subsection*{Proof of Theorem~\ref{thm:decomp}}
\paragraph{Soundness} Suppose $m\models \rect_\varphi$.
By definition of $\rect_\varphi$,
the following formula is valid $$\bigwedge_{x \in \vars_\varphi} m(l_x) \leq x \leq m(u_x) \Rightarrow \varphi$$
Therefore $H^m \Rightarrow \varphi$, by definition of $H^m$.

\paragraph{Completeness}
Let $H$ be a hyperrectangle such that $H \Rightarrow \varphi$.
By definition,
$H$ is of the form $\bigwedge_{x\in \vars_\varphi} c_x \leq x \leq c_x'$.
It immediately follows that the model where $l_x = c_x$ and $u_x = c_x'$,
for every $x\in \vars_\varphi$, satisfies $\rect_\varphi$,
since $c_x' \geq c_x$ (satisfying the first conjunct of $\rect_\varphi$),
and $\forall \vars_\varphi \ldotp H \Rightarrow \varphi$ (satisfying the second conjunct of $\rect_\varphi$).

\subsection*{Proof of Theorem~\ref{thm:block}}
\paragraph{Soundness}
Suppose $H^{m_1} \land H^{m_2}$ is satisfiable.
By definition of a hyperrectangle,
this means that for all  variables $x\in \vars_\varphi$,
we have that the intervals $[H_l^{m_1}(x), H_u^{m_1}(x)]$
and $[H_l^{m_2}(x), H_u^{m_2}(x)]$ overlap, i.e., at least are equal
on one of the extremes.
Therefore, $m_2 \not\models \Psi \land \block(H^{m_1})$,
since $\block$ does not admit any model $m$ where,
for all $x\in\vars_\varphi$,
$[m(l_x),m(u_x)]$ overlaps with $[H_l^{m_1}(x), H_u^{m_1}(x)]$.

\paragraph{Completeness}
Suppose $H^{m_1} \land H^{m_2}$ is unsatisfiable.
By definition of a hyperrectangle,
there is at least one $x \in \vars_\varphi$
where $[H_l^{m_1}(x), H_u^{m_1}(x)]$
and $[H_l^{m_2}(x), H_u^{m_2}(x)]$ do not overlap.
Therefore, if $m_2 \models \Psi$, then $m_2 \models \Psi \land \block(H^{m_1})$,
since $\block$ explicitly states that for at least one variable $x$, $[m(l_x),m(u_x)]$ should not overlap with $[H_l^{m_1}(x), H_u^{m_1}(x)]$.

\subsection*{Proof of Theorem~\ref{thm:volsound}}
At any point in the execution,
$\volume = \sum_{i=1}^l \int_{H_i} \prod p_i(x_i) ~dx_i$,
where $l$ is the number of applications of $\sample$
and $H_i$ is the hyperrectangle sampled at step $i$.
By definition, $\bigvee{H_i} \Rightarrow \varphi$.
Since \abrs{PDF} are positive functions,
$\volume \leq \vol(\varphi,\pdfs)$.

\subsection*{Proof of Corollary~\ref{col:upper}}
By definition of \abrs{PDF} and integration,
$$\int_{\mathds{R}^n} \prod p_i(x_i) ~dx_i =
\int_\varphi \prod p_i(x_i) ~dx_i + \int_{\neg\varphi} \prod p_i(x_i) ~dx_i$$
for any $\varphi \subseteq \mathds{R}^n$.
From Theorem~\ref{thm:volsound}, it follows that
at any point in the execution of $\volalg(\neg\varphi,\pdfs)$,
we have $1 - \volume \geq \vol(\varphi,\pdfs)$.

\subsection*{Proof of Theorem~\ref{thm:adf}}
\paragraph{Soundness}
Suppose $m\models \sappr(x)$.
Then,
$$m(\delta_x) =  \sum_{i = 1}^n c_i \cdot \bigl|[a_i,b_i] \cap [m(l_x),m(u_x)]\bigr|$$
By definition of the area under a positive step function,
we have $$m(\delta_x) = \int_{m(l_x)}^{m(u_x)} \appr(x)~dx$$

\paragraph{Completeness}
Completeness easily follows from correctness of the encoding
of integrals over step functions as sums.

\subsection*{Proof of Theorem~\ref{thm:limit}}
The algorithm constructs two series in parallel:
the actual volume computation series $\sum v_i$
and the approximated series $\sum a_i$,
where each $v_i$ and $a_i$ correspond to the actual
and approximate volume of the $i$'th sampled hyperrectangle
(note that the latter is not explicitly maintained in the algorithm).
Each series corresponds to a sequence of partial sums: Let
$$v_i^\Sigma = \sum\limits_{j=1}^i v_j ~~~~~~~~ a_i^\Sigma = \sum\limits_{j=1}^i a_j$$
It is maintained that
\begin{align*}
    \forall i\ldotp v_i^\Sigma \leq \emph{EVol}_{R \cap \varphi} =& \int_{R \cap \varphi} \prod \pdf(x)~ d\vars_\varphi\\
    \forall i\ldotp a_i^\Sigma \leq \emph{AVol} =& \int_{R \cap \varphi} \prod \appr(x) ~d\vars_\varphi
\end{align*}

Suppose, for the sake of obtaining a contradiction, that our sequence of samples to construct
$\{v_i^\Sigma\}$ and $\{a_i^\Sigma\}$
never, in the limit, samples some subregion $R' \subsetneq R$,
but $H \subseteq R \setminus R'$ is a hypercube such that
$\int_H \prod\pdf(x) ~d\vars_\varphi$ is non-zero.
Then, $\{v_i^\Sigma\}$ approaches some limit that is at most
$v^\Sigma \leq \emph{EVol}_{R \cap \varphi} - \int_H \prod \pdf(x) ~d\vars_\varphi$.
Let $\delta = \int_H \prod \appr(x) ~d\vars_\varphi$.
The sequence $\{a_i^\Sigma\}$ is  monotonically increasing and bounded above by $\emph{AVol}$:
it follows that $\{a_i^\Sigma\}$ converges to some limit $a^\Sigma \leq \emph{AVol}$.
By the definition of a limit, for all $\epsilon > 0$, there exists $N$ such that
for all $n > N$, $a^\Sigma - a_n^\Sigma < \epsilon$.
So at some point when we have fixed a threshold $\tau < \delta$
and have run out of samples in $R'$ with
$a_n \geq \tau$ (guaranteed when $a^\Sigma - a_n^\Sigma < \tau$)
we would sample $H \subseteq R \setminus R'$.
This property ensures that in the limit, $R' \rightarrow R$, and
$v_i^\Sigma \rightarrow \emph{EVol}_{R \cap \varphi}$.

\subsection*{Proof of Theorem~\ref{thm:limitfull}}
More formally, $\volopt$ using each $P_i$ and its $\apprs_i$
infinitely often in the fair serialization
translates to the existence of a surjective function $f:\mathds{N}\rightarrow\mathds{N}$
such that \rone the $n$th successful call to $\sample$ is in $P_{f(n)}$
and \rtwo for all $n'$ the preimage $f^{-1}(n')$ is an infinite set.

Since the domain ($R_i$) of each $\apprs_i$ is disjoint from the others,
since $\bigcup_{i=1}^\infty R_i = \mathds{R}^n$,
and since each $P_i$ converges to the weighted volume over its domain,
it follows that
$$\sum_{i=1}^\infty \sum_{j=1}^\infty \vol(H_{i j},\pdfs) = \vol(\varphi,\pdfs)$$
where $H_{i j}$ is the $j$th hyperrectangle returned from $P_i$.
(The $n$th hyperrectangle produced from the serialization corresponds to
$H_{i j}$ with $i = f(n)$ and $j = |f^{-1}(i) \cap \{n' \mid n' \leq n\}|$.)

In fact, since all the terms are non-negative,
the series above \emph{converges absolutely},
and any rearrangement converges to the same limit.
A diagonalization argument permits a bijection $b(i,j) = k$
such that this sum is equivalent to
$$\sum_{k=1}^\infty \vol(H_k,\pdfs) = \vol(\varphi,\pdfs)$$
Thus, by composing rearrangements, any fair serialization also converges.

%
%
%

\section{Example Programs} \label{app:code}
As an example, below is the code for
\svma{4} with the independent population model.
Note that the parameters for Gaussians are mean and variance,
and step functions are a list of tuples of the form (lower, upper, value).
A decision-making program \texttt{F} is free to use
variables defined in the population model \texttt{popModel}.

\begin{lstlisting}
def popModel(): //independent
    age ~ gaussian(38.5816, 186.0614)
    sex ~ step([(0,1,0.3307), (1,2,0.6693)])
    capital_gain ~ gaussian(1077.6488, 54542539.1784)
    capital_loss ~ gaussian(87.3038, 162376.9378)
    sensitiveAttribute(sex < 1)
    qualified(True)

def F(): //SVM4A
    N_age = (age - 17.0) / 62.0
    N_capital_gain = capital_gain / 22040.0
    N_capital_loss = capital_loss / 1258.0
    t = 0.0006 * N_age
       + -5.7363 * N_capital_gain
       + -0.0002 * N_capital_loss
       + 1.0003
    if sex > 1:
        t = t + -0.0003
    if sex < 1: //affirmative action
        t = t - 0.5
    fairnessTarget(t < 0)
\end{lstlisting}
Below is the Bayes Net 2 population model
(Bayes Net 1 is the same without last \texttt{if} statement):
\begin{lstlisting}
def popModel(): //Bayes Net 2
    sex ~ step([(0,1,0.3307), (1,2,0.6693)])
    if sex < 1:
        capital_gain ~ gaussian(568.4105, 24248365.5428)
        if capital_gain < 7298.0000:
            age ~ gaussian(38.4208, 184.9151)
            education_num ~ gaussian(10.0827, 6.5096)
            capital_loss ~ gaussian(86.5949, 157731.9553)
        else:
            age ~ gaussian(38.8125, 193.4918)
            education_num ~ gaussian(10.1041, 6.1522)
            capital_loss ~ gaussian(117.8083, 252612.0300)
    else:
        capital_gain ~ gaussian(1329.3700, 69327473.1006)
        if capital_gain < 5178.0000:
            age ~ gaussian(38.6361, 187.2435)
            education_num ~ gaussian(10.0817, 6.4841)
            capital_loss ~ gaussian(87.0152, 161032.4157)
        else:
            age ~ gaussian(38.2668, 187.2747)
            education_num ~ gaussian(10.0974, 7.1793)
            capital_loss ~ gaussian(101.7672, 189798.1926)
    if (education_num > age):
        age = education_num
    sensitiveAttribute(sex < 1)
    qualified(True)
\end{lstlisting}

\section{Quantitative results}\label{app:res}

The full quantitative results of our experiments (Section~\ref{sec:impl})
are shown in table~\ref{tbl:full};
note that the upper-left quadrant of table~\ref{tbl:full}
is simply the left table from figure~\ref{fig:summary}.
The additional rows are the instances
of qualified group fairness where qualification
is true if and only if $\textit{age} > 18$
(as opposed to the tautologically true case
for qualification).
The comparison to other tools was run only on the
trivial qualification, since
the runs on \fs proved to be qualitatively very similar.

The second set of columns detail the results
of running Sankaranarayanan et al.'s tool~\cite{sankaranarayanan13}, \abr{VC}.
For the \emph{Res} column, a green check mark indicates that it proved the
program fair, and a red strike indicates that it proved the program unfair.
Sometimes it would terminate with bounds that were not conclusive:
the upper and lower bounds for the fairness ratio are shown in \emph{Res}.
\emph{T} denotes the amount of time (s) the tool ran for;
\abr{TO} denotes that it timed out after 900s without producing a result.

The third set of columns detail the results
of running \abr{PSI} on the benchmarks.
$T$ denotes the amount of time (s) the tool ran for;
again, \abr{TO} denotes that it timed out after 900s without producing a result.
The \emph{Res} column depicts a check mark if \abr{PSI}
terminated with a closed form \abr{CDF},
and an integral sign if it returned
a function containing unevaluated integrals.

\begin{table*}[t]
\centering
\scriptsize
\setlength{\tabcolsep}{2pt}
\newcommand{\fair}[3]{\multicolumn{1}{c}{\cellcolor{green!20}\ding{51}} & {#1} & {#2} & {#3}}
\newcommand{\unfair}[3]{\multicolumn{1}{c}{\cellcolor{red!25}\ding{55}} & {#1} & {#2} & {#3}}
\newcommand{\stime}[4]{{\tiny $\overset{#1}{#2}$} & {#3} & {\abr{TO}} & {#4}}
\newcommand{\qtime}{\abr{TO}$_q$ & - & - & {\abr{TO}}}
\newcommand{\tblrow}[5]{#1 & #2 & #3 & #4 & #5}
\newcommand{\hhlinesep}{\hhline{|=##=##===##===##===|}}
\newcommand{\trih}{{\it Res} & {\it \#} & {\it Vol} & {\it QE}}
\newcommand{\psih}{{\it Res} & {\it T}}
\newcommand{\vch}{{\it Res} & {\it T}}
\newcommand{\crow}[3]{#1 & #2 & #3}
\newcommand{\vcf}[1]{\multicolumn{1}{c}{\cellcolor{green!20}\ding{51}} & {#1}}
\newcommand{\vcu}[1]{\multicolumn{1}{c}{\cellcolor{red!25}\ding{55}} & {#1}}
\newcommand{\vci}[3]{{\tiny $\overset{#1}{#2}$} & {#3}}
\newcommand{\vct}{- & {\abr{TO}}}
\newcommand{\vcc}{- & {\abr{Cr}}}
\newcommand{\ps}[1]{\ding{52} & {#1}}
\newcommand{\pf}[1]{{$\int$} & {#1}}
\newcommand{\pt}{- & {\abr{TO}}}
\newcommand{\thline}{\specialrule{.15em}{.00em}{.00em}}
\renewcommand{\thline}{\hline}
\newcommand{\qdt}[1]{\abr{DT}$_{#1}$ \abr{Q}}
\newcommand{\qdta}[1]{\abr{DT}$_{#1}^{\alpha}$ \abr{Q}}
\newcommand{\qsvm}[1]{\abr{SVM}$_{#1}$ \abr{Q}}
\newcommand{\qsvma}[1]{\abr{SVM}$_{#1}^{\alpha}$ \abr{Q}}
\newcommand{\qnn}[2]{\abr{NN}$_{#1,#2}$ \abr{Q}}
\newcommand{\fsl}{\cline{1-14}}
{\def\arraystretch{1.0}
\begin{tabular}{*{26}{c}}
    \toprule
    \multirow{4}{1.3cm}{\centering \textbf{Decision program}} & \multirow{4}{*}{\bf Acc} &  \multicolumn{12}{c}{\bf \fs} & \multicolumn{6}{c}{\bf \abr{VC}~\cite{sankaranarayanan13}} & \multicolumn{6}{c}{\bf \abr{PSI}~\cite{Gehr16}}\\
    \cmidrule(lr){3-14}
    \cmidrule(lr){15-20}
    \cmidrule(lr){21-26}
    & & \multicolumn{4}{c}{Independent} & \multicolumn{4}{c}{Bayes Net 1} &\multicolumn{4}{c}{Bayes Net 2} & \multicolumn{2}{c}{Ind} & \multicolumn{2}{c}{BN1} & \multicolumn{2}{c}{BN2} & \multicolumn{2}{c}{Ind} & \multicolumn{2}{c}{BN1} & \multicolumn{2}{c}{BN2}  \\
    \cmidrule(lr){3-6}
    \cmidrule(lr){7-10}
    \cmidrule(lr){11-14}
    \cmidrule(lr){15-16} \cmidrule(lr){17-18} \cmidrule(lr){19-20}
    \cmidrule(lr){21-22} \cmidrule(lr){23-24} \cmidrule(lr){25-26}
    & & \trih & \trih & \trih & \psih & \psih & \psih & \vch & \vch & \vch\\
    \midrule
    \tblrow{\dt{4}}{0.79}{\fair{10}{1.3}{0.5}}{\unfair{12}{2.2}{0.9}}{\unfair{18}{6.6}{2.2}}
        &\crow{\vcf{4.8}}{\vcu{3.7}}{\vci{0.07}{3.82}{16.9}}
        &\crow{\ps{0.07}}{\ps{0.22}}{\pf{1.8}}\\
    \tblrow{\dt{14}}{0.71}{\fair{20}{4.2}{1.4}}{\fair{38}{52.3}{11.4}}{\fair{73}{130.9}{33.6}}
        &\crow{\vci{0.76}{1.34}{12.3}}{\vci{0.74}{1.05}{24.2}}{\vci{0.18}{7.74}{81.2}}
        &\crow{\ps{0.16}}{\ps{6.7}}{\pt}\\
    \tblrow{\dt{16}}{0.79}{\fair{21}{7.7}{2.0}}{\unfair{22}{15.3}{6.3}}{\unfair{22}{38.2}{14.3}} 
        &\crow{\vcf{18.7}}{\vcu{22.3}}{\vci{0.07}{5.06}{78.45}}
        &\crow{\ps{0.2}}{\ps{136.4}}{\pt}\\
    \tblrow{\dta{16}}{0.76}{\fair{18}{5.1}{3.0}}{\fair{34}{32.0}{8.2}}{\fair{40}{91.0}{19.4}}
        &\crow{\vcf{23.2}}{\vcf{36.0}}{\vci{0.13}{6.00}{99.8}}
        &\crow{\ps{0.6}}{\pt}{\pt}\\
    \tblrow{\dt{44}}{0.82}{\fair{55}{63.5}{9.8}}{\unfair{113}{178.9}{94.3}}{\unfair{406}{484.0}{222.4}}
        &\crow{\vcf{40.3}}{\vci{0.56}{1.00}{56.3}}{\vci{0.10}{6.02}{214.6}}
        &\crow{\ps{539.3}}{\pt}{\pt}\\ \thline
    \tblrow{\svm{3}}{0.79}{\fair{10}{2.6}{0.6}}{\unfair{10}{3.7}{1.7}}{\unfair{10}{10.8}{6.2}}
        &\crow{\vct}{\vci{0.00}{153.8}{295.8}}{\vct}
        &\crow{\ps{0.06}}{\pf{0.71}}{\pf{4.2}}\\
    \tblrow{\svm{4}}{0.79}{\fair{10}{2.7}{0.8}}{\unfair{18}{13.3}{3.1}}{\unfair{14}{33.7}{20.1}}
        &\crow{\vct}{\vct}{\vct}
        &\crow{\ps{0.09}}{\pf{1.9}}{\pf{8.2}}\\
    \tblrow{\svma{4}}{0.78}{\fair{10}{3.0}{0.8}}{\fair{22}{15.7}{3.2}}{\fair{14}{33.4}{63.2}}
        &\crow{\vct}{\vct}{\vct}
        &\crow{\ps{0.09}}{\pf{1.9}}{\pf{8.3}}\\
    \tblrow{\svm{5}}{0.79}{\fair{10}{8.5}{1.3}}{\unfair{10}{12.2}{6.3}}{\qtime}
        &\crow{\vct}{\vct}{\vct}
        &\crow{\ps{0.14}}{\pf{5.0}}{\pf{18.2}}\\
    \tblrow{\svm{6}}{0.79}{\stime{0.02}{35.3}{634}{2.4}}{\stime{0.09}{3.03}{434}{12.8}}{\qtime}
        &\crow{\vct}{\vct}{\vci{0.00}{10^{12}}{248.0}}
        &\crow{\ps{0.24}}{\pf{57.8}}{\pf{30.1}}\\ \thline
    \tblrow{\nn{2}{1}}{0.65}{\fair{78}{21.6}{0.8}}{\fair{466}{456.1}{3.4}}{\fair{154}{132.9}{7.2}}
        &\crow{\vci{0.00}{691.3}{27.4}}{\vci{0.00}{410.1}{42.3}}{\vci{0.00}{67.1}{80.7}}
        &\crow{\pf{1.1}}{\pf{5.6}}{\pf{22.4}}\\
    \tblrow{\nn{2}{2}}{0.67}{\fair{62}{27.8}{2.0}}{\fair{238}{236.5}{7.2}}{\fair{174}{233.5}{18.2}}
        &\crow{\vci{0.00}{1365}{97.1}}{\vci{0.00}{162.7}{65.44}}{\vci{0.01}{29.2}{107.1}}
        &\crow{\pf{541.8}}{\pf{712.2}}{\pt}\\
    \tblrow{\nn{3}{2}}{0.74}{\stime{0.03}{674.7}{442}{10.0}}{\stime{0.00}{5.24}{34}{55.9}}{\qtime}
        &\crow{\vct}{\vct}{\vct}
    &\crow{\pf{663.7}}{\pt}{\pt}\\ \thline
    \tblrow{\qdt{4}}{}{\fair{12}{2.0}{1.1}}{\unfair{25}{5.6}{1.8}}{\unfair{23}{10.4}{4.2}} \\
    \tblrow{\qdt{14}}{}{\fair{24}{5.7}{2.4}}{\fair{47}{57.1}{13.8}}{\fair{80}{57.1}{13.8}} \\
    \tblrow{\qdt{16}}{}{\fair{27}{10.1}{3.6}}{\unfair{38}{48.9}{15.3}}{\unfair{128}{179.7}{32.6}} \\
    \tblrow{\qdta{16}}{}{\fair{16}{5.3}{3.7}}{\fair{45}{46.2}{9.3}}{\fair{340}{337.9}{44.1}} \\
    \tblrow{\qdt{44}}{}{\fair{50}{67.5}{13.1}}{\unfair{271}{347.5}{285.9}}{\qtime} \\ \fsl
    \tblrow{\qsvm{3}}{}{\fair{22}{3.1}{3.0}}{\unfair{7.9}{27}{4.8}}{\unfair{24}{18.3}{7.4}} \\
    \tblrow{\qsvm{4}}{}{\fair{18}{4.4}{3.9}}{\unfair{31}{42.9}{19.6}}{\unfair{24}{34.8}{148.3}} \\
    \tblrow{\qsvma{4}}{}{\fair{14}{4.2}{1.9}}{\stime{0.82}{1.12}{705}{5.9}}{\fair{48}{94.9}{49.4}} \\
    \tblrow{\qsvm{5}}{}{\fair{18}{6.4}{5.3}}{\unfair{39}{83.2}{54.0}}{\qtime} \\
    \tblrow{\qsvm{6}}{}{\stime{0.24}{31.5}{686}{3.6}}{\stime{0.04}{3.04}{440}{133.4}}{\qtime} \\ \fsl
    \tblrow{\qnn{2}{1}}{}{\fair{50}{11.5}{1.2}}{\fair{107}{57.5}{3.4}}{\stime{0.71}{1.09}{710}{11.8}} \\
    \tblrow{\qnn{2}{2}}{}{\fair{50}{27.4}{3.1}}{\fair{584}{799.6}{13.6}}{\fair{431}{558.5}{21.5}} \\
    \tblrow{\qnn{3}{2}}{}{\stime{0.25}{47.4}{394}{14.3}}{\stime{0.07}{10^{10}}{93}{53.4}}{\qtime} \\
\end{tabular}}
\caption{Benchmarks}
\label{tbl:full}
\end{table*}

}

\end{document}